\documentclass[prd,amsfonts,twocolumn,nofootinbib,showpacs]{revtex4}
\RequirePackage{graphicx}
\usepackage{enumerate}
\usepackage{bm}
\usepackage{braket}
\usepackage{slashed}
\usepackage{comment}
\begin{document}
\title{Baryogenesis from the Berry phase}

\author{Seishi Enomoto}
\affiliation{Department of Physics, University of Florida, Gainesville, Florida 32611, USA}
\affiliation{Theory Center, High Energy Accelerator Research Organization (KEK),Tsukuba, Ibaraki 305-0801, Japan}
\author{Tomohiro Matsuda}
\affiliation{Laboratory of Physics, Saitama Institute of Technology,
Fukaya, Saitama 369-0293, Japan}
\begin{abstract}
The spontaneous baryogenesis scenario explains how a baryon asymmetry
 can develop while baryon violating interactions are still in thermal
 equilibrium.
However, generation of the chemical potential from the derivative
 coupling is dubious since the chemical potential may not appear after
 the Legendre transformation.
The geometric phase (Pancharatnam-Berry phase) results from the
 geometrical properties of the parameter space of the Hamiltonian, which
 is calculated from the Berry connection.
In this paper, using the formalism of the Berry phase, we show that the
 chemical potential defined by the Berry connection is consistent
 with the Legendre transformation.
The framework of the Berry phase is useful in explaining the
 mathematical background of the spontaneous baryogenesis and also in
 calculating the asymmetry of the nonthermal particle
 production in time-dependent backgrounds.
Using the formalism, we show that the mechanism can be extended to more
 complex situations.
\end{abstract}

\maketitle
\section{Introduction}
Quantum mechanics is distinguishable from the classical counterpart 
by the phase factor, which explains
many characteristic phenomena of the quantum theory.
Among those, the Aharonov-Bohm(AB) effect\cite{Aharonov:1959fk} illuminates
the importance of the geometric phase in quantum mechanics.
It explains why an interference pattern can appear even though a magnetic
field is confined in a solenoid and put away from the orbit.
The phase originating from the geometry is called the Pancharatnam-Berry phase
or the Berry phase in short\cite{Berry:1984jv, Pancharatnam:1956}.
Suppose that the normalized state $\ket{\psi(t)} \in {\cal H}$
obeys the Schr\"odinger equation\cite{Aharonov:1987gg},
\begin{eqnarray}
i\hbar\frac{d}{dt}\ket{\psi(t)}&=&H(t)\ket{\psi(t)},
\end{eqnarray}
where $\ket{\psi(\tau)}=e^{i\phi} \ket{\psi(0)}$ for an interval $[0,\tau]$.
If we define $\ket{\tilde{\psi}(t)}=e^{-if(t)}\ket{\psi(t)}$ such that
$f(\tau)-f(0)=\phi$, we find
$\ket{\tilde{\psi}(t)}=\ket{\tilde{\psi}(0)}$, but the Schr\"odinger
equation for the new field becomes
\begin{eqnarray}
i\hbar\frac{d}{dt}\ket{\tilde{\psi}(t)}&=&H(t)\ket{\tilde{\psi}(t)}
+\hbar\frac{df}{dt}\ket{\tilde{\psi}(t)},
\end{eqnarray}
where the last term gives the Berry connection.
To understand the nonadiabatic contribution from the state mixing,
consider a slowly varying $H(t)$ with $H(t)\ket{n(t)}=E_n(t)\ket{n(t)}$
and write
\begin{eqnarray}
\ket{\psi(t)}&=&\sum_n a_n(t)e^{-\frac{i}{\hbar}\int E_n dt}\ket{n(t)}.
\end{eqnarray}
Then we have 
\begin{eqnarray}
\dot{a}_m&=&-a_m\bra{m}\frac{d}{dt}\ket{m}\nonumber\\
&&-\sum_{n\ne m}a_n
 \frac{\bra{m}\dot{H}\ket{n}}{E_n-E_m}e^{-\frac{i}{\hbar}\int (E_m-E_n)
 dt},
\end{eqnarray}
where the second term is negligible (by definition) 
in the adiabatic limit, since the
adiabatic limit is defined for the evolution without transition between states.
The phase coming from the first term is the conventional Berry phase,
which may appear both in the adiabatic and the nonadiabatic evolutions.
If the phase appears from the state mixing, it is called the
nonadiabatic Berry phase.
In contrast to the conventional Berry phase, the nonadiabatic Berry
phase does not appear in the adiabatic limit.
We hope there is no confusion between the ``Berry phase in a nonadiabatic
evolution'' and ``the nonadiabatic Berry phase''. They have
different origins.\footnote{See also Ref.\cite{Kayanuma:1994}.}

As we explain later, the first term (the Berry connection) gives the
chemical potential when the spontaneous baryogenesis scenario is
considered in the formalism
of the Berry phase.
However, since the Berry connection vanishes in the adiabatic limit
(although its integral may not vanish in a topological background),
 the evolution has to be
nonadiabatic in order to generate a sensible chemical potential. 
When we consider the spontaneous baryogenesis scenario, the second term (or the higher
terms) gives the particle production due to the time-dependent 
background.

To show our idea in a simple model, we start with the Schr\"odinger
equation for the state $\psi_0^t\equiv 
(K^0,\bar{K}^0)$, which is written as\footnote{
Our discussion here is implicitly based on a kaon, where $K^0$ is a
neutrally charged scalar meson.
We consider the model since the kaon is the simplest and the most familiar
among particle physicists. 
Note however that our baryogenesis scenarios are not for the kaon
production.
The idea will be applied to more complex scenarios.}
\begin{eqnarray}
\label{eq-schrodinger}
i\frac{d}{dt}\psi_0&=&H\psi_0,\nonumber\\
\left(
\begin{array}{cc}
H_{11} & H_{12}\\
H_{21} & H_{21}
\end{array}
\right)&=& 
\left(
\begin{array}{cc}
M & \Delta\\
\Delta^* & M
\end{array}
\right).\label{eq:schrodinger}
\end{eqnarray}
Here $K^0$ and $\bar{K}^0$ represent the matter and the antimatter
states of a singlet, and $\Delta\equiv |\Delta|e^{i\theta}$.
As far as the parameters are both homogeneous in space and static in
time, one can always find the Hamiltonian with real $\Delta$
($\theta=0$), using the rotation of the states.
In that case, the effective theory does not depend explicitly on
$\theta$, and the Hamiltonian is given by\footnote{Here, the capital ``R'' is
for the real off-diagonal elements and ``E'' is for the eigenstates.}
\begin{eqnarray}
H^R=\left(
\begin{array}{cc}
M & |\Delta|\\
|\Delta| & M
\end{array}
\right).
\end{eqnarray}
The rotation (redefinition) of a field is commonly used in removing
phase factors in the theory. 
Here, such a ``trivial'' transformation is being used in a
time-dependent background.
One may claim that this is a gauge transformation without the
gauge symmetry.

If $\theta$ is time dependent, one cannot neglect the time dependence of
the transformation matrix.
The rotation can be written using the unitary matrix $U_\theta$, which we
define
\begin{eqnarray}
\label{eq-theta}
\psi^R&\equiv& U_\theta^{-1} \psi_0,\nonumber\\
U_\theta&\equiv& 
\left(
\begin{array}{cc}
e^{i\theta(t)/2} & 0\\
0 & e^{-i\theta(t)/2}
\end{array}
\right).
\end{eqnarray}
Then, the Schr\"odinger equation for the state $\psi^R$ is written as
\begin{eqnarray}
i\frac{d}{dt}\psi^R&=&\left(H^R-iU_\theta^{-1}\dot{U}_\theta\right)\psi^R. 
\end{eqnarray}
The original (``trivial'') transformation is a global transformation and
gives nothing from the left-hand side.
On the other hand, since we have introduced the time dependence, the
transformation is a local transformation and gives the additional
contribution from the time derivative, which is called the Berry
connection.

Note that $\psi^R$ is not the eigenstate of the Hamiltonian.
In this model, the eigenstate can be written as
\begin{eqnarray}
\psi^E&=&U_1^{-1}\psi^R =U_1^{-1}U_\theta^{-1}\psi_0,
\end{eqnarray}
where 
\begin{eqnarray}
U_1&\equiv& 
\frac{1}{\sqrt{2}}\left(
\begin{array}{cc}
1 & 1\\
-1 & 1
\end{array}
\right).
\end{eqnarray}
The eigenstate $\psi^E$ is the true eigenstate of the Hamiltonian {\bf only 
when} $U_1^{-1}U_\theta^{-1}$ is not time dependent.
Therefore, we sometimes denote $\psi^E$ with the double
quotation marks (``eigenstate'') in the time-dependent
background.\footnote{See also Appendix \ref{app-bias}}

Formally, the equivalence class of state vectors or ``projective Hilbert
space'' is defined using an arbitrary function $U$ as
$\{U^{-1}\psi_0\}$, and an equivalence 
class of Hamiltonians is $\{U^{-1}HU-iU^{-1}(\partial_t
U)\}$\cite{Aharonov:1987gg,Samuel:1988zz}\footnote{Here, $R$ in $U(R)$
represents arbitrary parameters.}.
These are defining different representations of the identical
Schr\"odinger equation. 
Note that the Berry connection depends on the choice of the state
vector.
Although $U^{-1}H U=H$ is true for the Abelian model, a non-Abelian
extension is possible, in which one has to consider $U^{-1}H U\ne H$.
In this case, the Hamiltonian is not invariant under the
transformation, but (therefore) it can be used to remove the phase
parameter of the Hamiltonian during the
time-dependent background. 

The Berry phase is defined by the integral of the Berry connection
along the orbit. 
If the Berry phase is defined for a cyclic process
starting from $t=0$ and ends at $t=T$, the Hamiltonian at $t=0$
and $t=T$ must coincide. 
Since the process considered in this paper is not a cyclic process, the
definition of the Berry connection can be ambiguous. To avoid such
ambiguity, we are always choosing the state, which removes the time-dependent
phase in the Hamiltonian.
In the above argument, instead of considering the cyclic
process, $U$ has been chosen to keep the phase parameter
of $H^R$ unchanged along the classical orbit.
In this paper, we sometimes call this specific transformation ``the Berry
transformation''.
This is not a common terminology since the Berry phase is usually
defined using the cyclic process.
In this paper, the Berry connection is defined using the transformation.

One will find that the mechanism is similar to the spontaneous
baryogenesis
scenario\cite{Cohen:1987vi,Cohen:1988kt,Cohen:1990it,Dine:1990fj}, 
in which the effective chemical potential is coming from the derivative
coupling of the Nambu-Goldstone boson, not from the Berry connection.
We will discuss the discrepancy in Sec. \ref{sec-legendre}.
Here, we have at least three reasons to consider 
the Berry phase in the spontaneous baryogenesis scenario.
The primary reason is the consistency between the Lagrangian
and the Hamiltonian formalisms. (See Sec. \ref{sec-legendre}.)
Second, the formalism based on the Berry phase is free from the
spontaneous symmetry breaking.
As we will see in Sec. \ref{sec:berrytochem}, the origin of the chemical
potential may not be the Nambu-Goldstone boson.
Therefore, there is a hope that baryogenesis with the
Berry phase is giving a natural extension of the scenario,
i.e, ``not a spontaneous'' baryogenesis in which there is no
Nambu-Goldstone boson and the symmetry is explicitly violated.
Third, using the formalism based on the Berry phase, one can 
see the mathematical structure of the model.
In addition to the conventional chemical potential, the
nonadiabatic Berry phase may appear.\footnote{See
Refs.\cite{Kayanuma:1994, Moore:1992, Oka:2009ek} for example.}.  

Normally, when one discusses the nonadiabatic effect for the Berry phase, 
his (her) motivation would be to calculate the Berry and the
nonadiabatic Berry phases. 
However, our present discussion is not for the calculation of the Berry
phase in a cyclic process, but for finding the sources of the
asymmetry in time-dependent backgrounds. 
We hope there is no misdirection in our arguments.

In the next section, using simple setups, we are going to discuss why the 
formalism of the Berry phase can be used to understand the scenario of
the spontaneous baryogenesis.
Then, we will consider some extension of the scenario, to solve more
complex situations.

\section{Effective chemical potential and the Berry phase}
\label{sec:berrytochem}
In the early Universe, a field can be placed away from its true minimum.
Then the field starts to roll down on the potential during the evolution of
the Universe, and it starts to oscillate around the minimum.
Sometimes, the trajectory of the oscillation is not a straight line passing
through the minimum, but an oval form, since a CP
violating interaction may introduce angular rotation of the
field\cite{Affleck-Dine}. 
One can also imagine that the symmetry breaking occurs at a high energy scale
and the effective action is written using a quasi Nambu-Goldstone boson (axion).
These are the standard realization of the field rotation.

Below, we are going to explain the basic idea 
using the kaonlike model.
From the Schr\"odinger equation (\ref{eq-schrodinger}), 
we find for $\psi^R$;
\begin{eqnarray}
i\frac{d}{dt}\psi^R&=&(H^R -iU_\theta^{-1}\dot{U}_\theta)\psi^R,\nonumber\\
H^R&=& 
\left(
\begin{array}{cc}
M & |\Delta|\\
|\Delta| & M
\end{array}
\right),\nonumber\\
iU_\theta^{-1}\dot{U}_\theta&=&
\frac{1}{2}\left(
\begin{array}{cc}
\dot{\theta} & 0\\
0 & -\dot{\theta}
\end{array}
\right),\label{eq:chemical_potential_schrodinger}
\end{eqnarray}
where $\mu\equiv \dot{\theta}/2$ can be regarded as an effective chemical
potential. 
Here, we defined $\theta$ as $\Delta\equiv |\Delta|e^{i\theta}$.
From these equations, the relation between the effective chemical
potential and the Berry connection is very clear.

Above, we have calculated the Berry connection with respect
to $\psi^R$, but one will soon find that $\psi^R$ is not the eigenstate
of the equation.
Of course, a similar discussion can be applied to the eigenstate, but the
appearance of the chemical potential is not obvious.
Below, we will show what happens if one chooses the eigenstate for the discussion. 

If $\theta$ does not depend on time, one can calculate the eigenstate,
which is given by
\begin{eqnarray}
\psi^E\equiv \left(\psi^{E(+)}, \psi^{E(-)}\right)^t,\nonumber\\
\psi^{E(\pm)}\equiv\left(\pm \frac{e^{i\theta/2}}{\sqrt{2}}K^0+\frac{e^{-i\theta/2}}{\sqrt{2}}\bar{K}^0\right).
\end{eqnarray}
On the other hand, if $\theta$ is time dependent, one has to introduce
the Berry connection to the Schr\"odinger equation, which becomes 
\begin{eqnarray}
i\frac{d}{dt}\psi^E&=&(H^E -iU^{-1}\dot{U})\psi^E,\nonumber\\
H^E&=& 
\left(
\begin{array}{cc}
M+|\Delta| & 0\\
0 & M-|\Delta|
\end{array}
\right),\nonumber\\
iU^{-1}\dot{U}&=&
\frac{1}{2}\left(
\begin{array}{cc}
0 & \dot{\theta}\\
\dot{\theta} &0 
\end{array}
\right),
\end{eqnarray}
Obviously, the ``eigenstate'' $\psi^{E}$ is no longer the
true eigenstate of the  Schr\"odinger equation because of the mixing
caused by the Berry connection.
Also, unlike the calculation based on $\psi^R$, it seems difficult to
understand that the Berry connection works like an effective chemical
potential.
This is the reason why we consider $\psi^R$ instead of using $\psi^E$.

In the past, the effective chemical potential has been studied in
particle cosmology in various ways.
Spontaneous baryogenesis scenario uses higher-dimensional operators such
as\cite{Cohen:1987vi,Cohen:1988kt,Cohen:1990it,Dine:1990fj} 
\begin{eqnarray}
\label{eq-eff}
{\cal O}_{h}=-\frac{\partial_{\mu} \varphi}{M_*}J^\mu_B,
\end{eqnarray}
where $\varphi$ is a scalar field and $J^\mu_B$ is the baryon current.
Similarly, one can calculate the effective chemical potential
using the effective Lagrangian for the Nambu-Goldstone
boson\cite{Dine:1981rt,Dolgov:1994zq}.  
On the other hand, in our case, since the Berry connection is defined
for the parameter on the classical orbit, the chemical potential has to
be defined for the parameter, not for the field. 
This point is crucial when we consider the Legendre transformation.
The underlying problem of the derivative coupling has been discussed by
Arbuzova et al. in Ref.\cite{Arbuzova:2016qfh} and by Dasgupta et al. in
Ref.\cite{Dasgupta:2018eha}. 
We will discuss this issue in Sec. \ref{sec-legendre}.

It is easy to show that the chemical potential in the Hamiltonian can
bias the particle number densities in the thermal equilibrium.
In that sense, the appearance of the chemical potential in the
Hamiltonian explains the asymmetry in the thermal equilibrium.

\subsection{The Berry transformation in the Lagrangian}
Before moving forward, it will be useful to show explicitly the relation
between the chemical potential and the Berry phase in the Lagrangian. 
We start with the Hamiltonian
\begin{equation}
 H= H_0-\mu J^0,
\end{equation}
where $J^0$ is the net number of particles.
Here, we consider $H_0$ as a simple Hamiltonian given by a complex scalar $\chi$
and its conjugate momentum $\pi^*\equiv\partial\mathcal{L}/\partial\dot{\chi}$ as
\begin{equation}
 H_0= \int d^3x\left(\pi^*\pi+\chi^*\omega^2\chi+V(\chi)\right) \label{eq:Hamiltonian0_chi}
\end{equation}
where $\omega^2\equiv-\nabla^2+m^2$ and $V$ is a potential.
On the other hand, since $J^0$ is $U(1)$ Noether charge, this is derived
from the original Lagrangian as
\begin{eqnarray}
 J^0 &=& \int d^3x \:i\left(\chi^*\frac{\partial\mathcal{L}}{\partial\dot{\chi}^*}
  -\frac{\partial\mathcal{L}}{\partial\dot{\chi}}\chi\right) \nonumber\\
  &=& \int d^3x \:i(\chi^*\pi-\pi^*\chi). \label{eq:J0_chi}
\end{eqnarray}
Using Eqs.(\ref{eq:Hamiltonian0_chi}), (\ref{eq:J0_chi}) and the Heisenberg equation
\begin{eqnarray}
 i\dot{\chi} &=& [\chi,H] \nonumber \\
  &=& i(\pi+i\mu\chi),
\end{eqnarray}
one can derive the original Lagrangian as
\begin{eqnarray}
 \int d^3x \mathcal{L} &=& \int d^3x (\pi^*\dot{\chi}
  +\dot{\chi}^*\pi)-H\nonumber \\
  &=& \int d^3x \left( |\dot{\chi}-i\mu\chi|^2-\chi^*\omega^2\chi-V(\chi) \right).
   \label{eq:Lagrangian0_chi1}
\end{eqnarray}
Therefore, the representation of the chemical potential in the Lagrangian
is similar to a gauge field $A_0$\footnote{
Equation (\ref{eq:Lagrangian0_chi1}) is also represented as
\begin{equation}
 \mathcal{L}=|\partial\chi|^2-(m^2-\mu^2)|\chi|^2-V +\mu J^0,
\end{equation}
where the original mass $m^2$ is replaced by $m^2-\mu^2$.
Such modification does not appear for the fermions.}.

Note that the replacement $\dot{\chi}\rightarrow\dot{\chi}-i\mu\chi$ is equivalent to
\begin{equation}
 \chi\rightarrow \tilde{\chi} \equiv\chi e^{-i\theta(t)}, \quad \theta(t)\equiv\int^t dt'\mu(t').
  \label{eq:chi_transformation}
\end{equation}
Applying this replacement to Eq.(\ref{eq:Lagrangian0_chi1}),
 the Lagrangian becomes 
\begin{equation}
 \mathcal{L}=|\partial\tilde{\chi}|^2-m^2|\tilde{\chi}|^2-V(\tilde{\chi} e^{i\theta}).
  \label{eq:chi_tilde_lagrangian}
\end{equation}
This Lagrangian does not have the effective chemical potential, but 
some interaction (e.g, $V \sim \chi^n + H.c.$) could not be invariant
under this replacement.
If the Lagrangian contains such interaction, the $e^{i\theta(t)}$ dependence will remain.
Note that in Eq.(\ref{eq:chemical_potential_schrodinger}), we obtained
the chemical potential using the unitary matrix $U_\theta$, which is
defined to remove the complex phases in the Hamiltonian.
In this sense, the inverse process (from Eq.(\ref{eq:chi_tilde_lagrangian})
to Eq.(\ref{eq:Lagrangian0_chi1})) is equivalent to the procedure from
Eq.(\ref{eq:schrodinger}) to Eq.(\ref{eq:chemical_potential_schrodinger}). 
In this respect, the phase $\theta(t)$ used above can be regarded as the Berry
phase, and also the chemical potential 
\begin{equation}
 \mu=i\left[e^{-i\theta}\right]^{-1}
  \cdot\partial_t\left[e^{-i\theta}\right]
\end{equation}
can be regarded as the Berry connection associated with the transformation
in Eq. (\ref{eq:chi_transformation}).

\subsection{Particle production with a time-dependent background}
\label{sec-production-scalar}
As a useful toy model, we first consider a time-dependent background for
a complex scalar field and calculate the perturbative particle
production, then examine the sources of the asymmetry.

We start with a complex scalar field $\chi$ with the time-dependent
mass\cite{Dolgov:1996qq}
\begin{eqnarray}
{\cal L}_{int}&=& F(t)\chi^* \chi.
\end{eqnarray}
We take $F(t)\rightarrow 0$ in the past and expand
\begin{eqnarray}
\chi=\int\frac{d^3p}{2\omega (2\pi)^3}\left[
a_{\mathbf{k}}e^{-i\omega t}
+b^\dagger_{-\mathbf{k}}e^{+i\omega t}
\right]e^{i\mathbf{k}\cdot \mathbf{x}}.
\end{eqnarray}
At later times, we define
\begin{eqnarray}
\chi=\int\frac{d^3p}{2\omega (2\pi)^3}\left[
a_{\mathbf{k}}f_k(t)
+b^\dagger_{-\mathbf{k}}g^*_{k}(t)
\right]e^{i\mathbf{k}\cdot \mathbf{x}},
\end{eqnarray}
which gives the equation of motion
\begin{eqnarray}
\left[
\partial_t^2+\mathbf{k^2}+m^2\right]f_k(t)=-F(t) f_k(t).
\end{eqnarray}
Expanding $f_k=f_k^0+f_k^1$ and $g_k=g_k^0+g_k^1$ 
for $f_k^0=e^{-i\omega t}$ and $g_k^0=e^{-i\omega t}$,
we find
\begin{eqnarray}
\left[
\partial_t^2+\mathbf{k^2}+m^2\right]f_k^1=-F(t) e^{-i\omega t}.
\end{eqnarray}
The conventional Green's function method gives
\begin{eqnarray}
f_k^1&=& \int\frac{d\omega'}{2\pi} \frac{\tilde{F}(\omega'-\omega)}
{\omega'^2-\omega^2}e^{-i\omega' t},
\end{eqnarray}
where $\omega'$ is coming from the time derivative of $f^1_k(t)$ and 
the Fourier transformations are defined as
\begin{eqnarray}
\tilde{F}(\omega)\equiv \int dt F(t) e^{i\omega t}\nonumber\\
\tilde{F}^*(\omega)\equiv \int dt F(t)^* e^{i\omega t}.
\end{eqnarray}
The pole at $\omega'=-\omega$ introduces $e^{+i\omega t}$ in $f_k^1$, which 
gives
\begin{eqnarray}
f_k&=& \alpha_k^f e^{-i\omega t}+\beta_k^{f*} e^{i\omega t},
\end{eqnarray}
where 
\begin{eqnarray}
\beta_k^{f}=i\frac{\tilde{F}(2\omega)^*}{2\omega}.
\end{eqnarray}
Considering the Bogoliubov transformation, one will find that
the particle and the antiparticle numbers are given by
\begin{eqnarray}
N_k&=&|\beta^f_k|^2 \nonumber\\
\bar{N}_k&=&|\beta^g_k|^2,
\end{eqnarray}
which explains why particles are produced in the time-dependent
background.
Obviously, in this case the source of the asymmetry is
\begin{eqnarray}
\left|\frac{\tilde{F}(-2\omega)^*}{2\omega}\right|^2\ne 
\left|\frac{\tilde{F}(-2\omega)}{2\omega}\right|^2,
\end{eqnarray}
which shows that the above model does not generate the asymmetry.

To introduce the bias, we introduce 
\begin{eqnarray}
{\cal L}_{int}&=& G(t)\chi \chi+G(t)^*\chi^* \chi^*.
\end{eqnarray}
Using the method given in Ref.\cite{Dolgov:1996qq}, one can calculate
the number densities from the amplitudes
\begin{eqnarray}
A&=&\bra{k_1, k_2}i\int d^4 x G(t)\chi \chi\ket{0}\nonumber\\
&=&i(2\pi)^3\delta^3(\mathbf{k}_1+\mathbf{k}_2)\int dt
 G(t)e^{i(\omega_1+\omega_2)}\nonumber\\
&=&i(2\pi)^3\delta^3(\mathbf{k}_1+\mathbf{k}_2) \tilde{G}(\omega_1+\omega_2)\\
\bar{A}&=&\bra{k_1, k_2}i\int d^4 x G(t)^*\chi^*
 \chi^*\ket{0}\nonumber\\
&=&i(2\pi)^3\delta^3(\mathbf{k}_1+\mathbf{k}_2) \tilde{G}^*(\omega_1+\omega_2),
 \end{eqnarray}
which gives 
\begin{eqnarray}
n&=&
 \int\frac{d^3k}{(2\pi)^3}\frac{|\tilde{G}(2\omega)|^2}{4\omega}\nonumber\\
\bar{n}&=&
 \int\frac{d^3k}{(2\pi)^3}\frac{|\tilde{G}(2\omega)^*|^2}{4\omega}.
\end{eqnarray}
In this case, $n-\bar{n}\ne 0$ is possible.
More specifically, the model discussed in Ref.\cite{Dolgov:1996qq} 
generates the interference between terms.
This is possible when multiple sources are introduced in $G(t)$.
The simplest example in this direction is given by
Eq.(\ref{eq-simp-expand}), which will be discussed in Sec. \ref{sec-decay}.

Let us see the origin of the asymmetry in the light of the chemical
potential and the Berry connection, not in terms of the interference
between terms.
To find the origin of the asymmetry, consider a constant (or a slowly
varying) chemical potential to define 
\begin{eqnarray}
G(t)=G_r e^{i \mu t},
\end{eqnarray}
where $G_r(t)$ is real.
Then, from the amplitudes we find
\begin{eqnarray}
A&=&i(2\pi)^3\delta^3(\mathbf{k}_1+\mathbf{k}_2)\int dt
 G_r (t)e^{i(\omega_1+\omega_2+\mu)}\nonumber\\
&=&i(2\pi)^3\delta^3(\mathbf{k}_1+\mathbf{k}_2) \tilde{G}_r(\omega_1+\omega_2+\mu)\\
\bar{A}&=&i(2\pi)^3\delta^3(\mathbf{k}_1+\mathbf{k}_2) \tilde{G}_r(\omega_1+\omega_2-\mu),
 \end{eqnarray}
In this case, the source of the asymmetry is
\begin{eqnarray}
\left|\frac{\tilde{G_r}(2\omega+\mu)}{2\omega}\right|^2\ne 
\left|\frac{\tilde{G_r}(2\omega-\mu)}{2\omega}\right|^2,
\end{eqnarray}
which is realized by $\mu\ne 0$.

Note that in the above case the nonadiabatic Berry phase may also appear 
since the particle production in the above argument is due to the
nonadiabatic transition between states.
The phase may not be important in the perturbative calculation discussed 
above, but it could be important in the nonperturbative
limit\cite{Kayanuma:1994}. 
In this paper, as a hint to understanding the topic, we will show an
interesting example in which the perturbative expansion 
does not show the asymmetry while the nonperturbative calculation
shows the asymmetry in the time-dependent background.
 (See Sec. \ref{sec-small-pert} and \ref{sec-majorana-simpleex}.)
This topic has to be studied in more detail using resurgent methods
\cite{Delabaere:1999, Dorigoni:2014hea}.

Normally, the Berry phase is not defined specifically for the
spontaneous violation of a symmetry.
A naive expectation is that the formalism based on the Berry phase may
be used in wider circumstances than the Nambu-Goldstone effective action.
To show how it works, we consider the simplest extension in the following,
to show that neither the spontaneous symmetry breaking nor the
derivative coupling is needed for generating the effective chemical
potential.
The model will be used also for the nonequilibrium particle production
in Sec. \ref{sec-decay}.\footnote{
The spontaneous baryogenesis can be discussed
      for (1) the chemical potential in the thermal equilibrium and (2)
      the nonperturbative particle creation caused by the
      time-dependent background. 
The latter can be discussed for the thermal equilibrium and may compensate the
simple discussion based on the chemical potential in the thermal background.
However, in our paper, we are considering the nonperturbative particle
      production only when the thermal background is negligible.
Therefore, we are calling the latter process ``nonequilibrium
      particle production'' and discriminate it from the former.
 }

\subsection{A small extension from the simple rotation}
\label{sec:exten}
\begin{figure}[t]
\centering
\includegraphics[width=1.0\columnwidth]{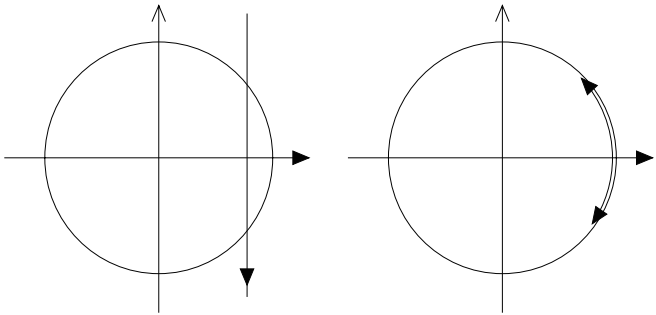}
 \caption{A straight motion with $m_R=\Lambda+ig\varphi$ is shown in the left,
 and the rotational oscillation with $m_R=|m_R|e^{i\theta(t)}$ is shown in the right.}
\label{fig-osc}
\end{figure}

We consider a simple example given by
\begin{eqnarray}
\Delta = \Lambda +ge^{i\hat{\theta}} \varphi,
\end{eqnarray}
where $\varphi$ is a ``real'' scalar field and $\Lambda, g$ are real
constant parameters.
$\hat{\theta}$ defines the direction of the motion.
Here, we consider a case with $e^{i\hat{\theta}}=i$ for simplicity.
Note that there is no $U(1)$ symmetry in this model, and the real field
$\varphi$ does not have a phase.
However, if one defines 
\begin{eqnarray}
\Delta\equiv |\Delta|e^{i\theta},
\end{eqnarray}
one can recover the argument of the Berry connection.
The chemical potential is calculated as
\begin{eqnarray}
\mu&=&\dot{\theta}=\frac{d}{dt}\arctan{\frac{g\varphi}{\Lambda}}\nonumber\\
&=&\frac{g\dot{\varphi}\Lambda}{\Lambda^2 +g^2\varphi^2}.
\end{eqnarray}
Figure \ref{fig-osc} shows the straight motion with
$\Delta(t)=\Lambda+ig\varphi(t)$ (in the left), which is compared with
the rotational oscillation with $\Delta(t)=|\Delta|e^{i\theta(t)}$ (in
the right). 
Note that we are {\bf not} using $\varphi$ in the derivative coupling.

\section{Particle production due to the background oscillations}
\label{sec-decay}
Our next topic is the nonequilibrium particle production in a more
realistic scenario. 
To compare our results with the conventional spontaneous baryogenesis,
we first review the calculation given in Ref.\cite{Dolgov:1996qq}.
They have considered the Lagrangian density
\begin{eqnarray}
\label{eq-Dolgov-L}
{\cal L}&=& \partial_\mu\Phi^*\partial^\mu\Phi -V(\Phi^*\Phi) \nonumber\\
&&+i\bar{Q}(i\gamma^\mu\partial_\mu-m_Q)Q+\bar{L}(i\gamma^\mu\partial_\mu-m_L)L\nonumber\\
&&+\left(g\Phi\bar{Q}L + H.c.\right),
\end{eqnarray}
which has the $U(1)$ symmetry corresponding to baryon number
\begin{eqnarray}
\label{eq-u1}
\Phi\rightarrow e^{i\alpha}\Phi, &Q\rightarrow e^{-i\alpha}Q,&
 L\rightarrow L.
\end{eqnarray}
Defining $\langle\Phi\rangle=fe^{i\theta}/\sqrt{2}$, one obtains an
effective Lagrangian density 
\begin{eqnarray}
{\cal L}&=& \frac{f^2}{2}\partial_\mu\theta \partial^\mu\theta \nonumber\\
&&+i\bar{Q}(i\gamma^\mu\partial_\mu-m_Q)Q+\bar{L}(i\gamma^\mu\partial_\mu-m_L)L\nonumber\\
&&+\left(\frac{ge^{i\theta}}{\sqrt{2}}f\bar{Q}L + H.c.\right).
\end{eqnarray}
$\theta$ in the above Lagrangian is the Nambu-Goldstone boson.
Considering the rotation 
\begin{eqnarray}
Q\rightarrow e^{-i\alpha}Q,& L\rightarrow L, &\theta\rightarrow \theta+\alpha
\end{eqnarray}
and assigning $\alpha=-\theta$, the Lagrangian gives\footnote{In our
formalisms of the Berry transformation, the assignment is
$\alpha=-\langle \theta \rangle$. 
Therefore, the chemical potential is replaced by
$\partial_\mu\langle\theta\rangle J^\mu$.
The difference is crucial for the Legendre transformation.}
\begin{eqnarray}
{\cal L}&=& \frac{f^2}{2}\partial_\mu\theta \partial^\mu\theta \nonumber\\
&&+i\bar{Q}(i\gamma^\mu\partial_\mu-m_Q)Q+\bar{L}(i\gamma^\mu\partial_\mu-m_L)L\nonumber\\
&&+\left(\frac{g}{\sqrt{2}}f\bar{Q}L + H.c.\right)+\partial_\mu\theta J^\mu,\label{eq:rotated_Lagrangian}
\end{eqnarray}
where $J^\mu=\bar{Q}\gamma^\mu Q$.

In the above, we have followed Ref.\cite{Dolgov:1996qq} and rotated
$Q$ to remove the phase.
However, one will soon find that the assignment of the rotation is not unique. 
Actually, if one rotates the fields as
\begin{eqnarray}
Q\rightarrow e^{i\alpha'(x)}Q,&& L\rightarrow e^{i\alpha'(x)}L,
\end{eqnarray}
the phases in the interaction remain the same.
On the other hand, one will find that
\begin{eqnarray}
\mathcal{L}_{\rm add}=-\partial_\mu\alpha'\cdot(\bar{Q}\gamma^\mu Q+\bar{L}\gamma^\mu L)
\label{eq:add_current}
\end{eqnarray}
appears.  
This term is related to the (global) $U(1)_{B+L}$
in the original Lagrangian.  Choosing $\partial_t\alpha'\neq0$, the 
chemical potential appears for the net $B+L$ number.
Therefore, if the task is just to remove the phase in the interaction,
the definition of the 
rotation (and of course the chemical potential) has the ambiguity.
This is not a surprise.
If one can use the relation $(n_Q-\bar{n}_Q)+(n_L-\bar{n}_{L})=0$, one
can always rewrite the chemical potential as
$\mu(n_Q-\bar{n}_{Q})\rightarrow\frac{\mu}{2}(n_Q-\bar{n}_{Q}) 
-\frac{\mu}{2}(n_L-\bar{n}_{L})$, where $\mu$ is $Q$'s chemical potential.
In the thermal equilibrium, the chemical potential has to balance.

Besides the symmetry discussed above, the Lagrangian is symmetric under the
exchange $\{L\leftrightarrow Q, \Phi\leftrightarrow \Phi^*\}$.
Assuming that the Berry transformation respects this symmetry, the
transformation has to be $Q\rightarrow e^{i\theta/2}Q$ and $ L\rightarrow
e^{-i\theta/2}L$ without ambiguity.\footnote{
Remember that in our previous discussion, $U_\theta$ in Eq.(\ref{eq-theta}) is unique 
because the rotation is defined for the matter and the antimatter.}

Using the calculation in Sec. \ref{sec-production-scalar}, we can
calculate the asymmetries, which appear both for $Q$ and $L$ with the
opposite signs.

The above arguments seem to be suggesting that the assignment 
$Q\rightarrow e^{i\theta/2}Q,\quad L\rightarrow
e^{-i\theta/2}L$ is more natural.
On the other hand, if one assumes that the coefficients are determined by
the $U(1)$ symmetry given by Eq.(\ref{eq-u1}), and claims that the chemical
potential is appearing form the derivative coupling of the
Nambu-Goldstone boson of the broken symmetry, the assignment 
seems to be unique.

Now consider the particle production in the time-dependent and
nonequilibrium background.
The production can be biased by the oscillation given by
\begin{eqnarray}
\Phi(t)&=&fe^{i\theta(t)},\nonumber\\
\theta(t)&=&\theta_ie^{-\Gamma t/2}\cos{m_\theta t}.
\end{eqnarray}
Expanding $\theta$ for small $\theta$ as
\begin{eqnarray}
\label{eq-simp-expand}
e^{i\theta(t)}=1+i\theta(t)-\theta(t)^2/2,
\end{eqnarray}
one will find\cite{Dolgov:1996qq} 
\begin{eqnarray} 
\Delta n_Q\equiv n_Q-\bar{n}_Q &=& \frac{g^2}{16\pi}m_\theta f^2\theta_i^3 \label{eq:delta_n_q}.
\end{eqnarray}
Note, however, that the above expansion of $e^{i\theta}$ already ruins
 the original symmetry.
Is the violation of the symmetry crucial for the asymmetry?
We will solve this problem in Sec. \ref{sec-majorana-osc} using the
 formalism of the Berry phase.

\subsection{Small extension and the perturbative expansion}
\label{sec-small-pert}
To check the validity of the above calculation in wider circumstances,
let us remove the condition 
\begin{eqnarray}
\Phi(t)&=&fe^{i\theta(t)},
\end{eqnarray}
and consider the interaction replaced by
\begin{eqnarray}
\label{eq-simplest}
g\Phi(t)&\rightarrow& \Lambda + g \varphi(t),
\end{eqnarray}
where $\Lambda$ is real but $g$ is a complex constant, and
$\varphi(t)$ is a time-dependent real scalar field.
Later, we will discuss the nonperturbative particle production, but in
this section we are confined to the perturbative expansion.
As is shown in Ref.\cite{Dolgov:1996qq}, the average number density $n$
of particle(or antiparticle) pairs produced by the decay of a homogeneous
classical scalar field can be calculated as 
\begin{eqnarray}
\Delta n_Q&=&\frac{1}{\pi^2}\int d\omega
 \omega^2\left|\int^\infty_{-\infty}
dt e^{2i\omega t}(\Lambda + g \varphi)\right|^2\nonumber\\
&&-\frac{1}{\pi^2}\int d\omega
 \omega^2\left|\int^\infty_{-\infty}
dt e^{2i\omega t}(\Lambda + g^* \varphi)\right|^2,
\end{eqnarray}
where $2\omega=p_a^0+p_b^0$ is for the particle pair $a$ and $b$.

Using $g=g_R+ig_I$, one can expand $(\Lambda + g \varphi)$ as
\begin{eqnarray}
(\Lambda + g \varphi)&=&\Lambda + ig_I \varphi + g_R\varphi.
\end{eqnarray}
The cross term that may give a nonzero contribution to the baryon
asymmetry is
\begin{eqnarray}
\Delta n_Q&=&\frac{2}{\pi^2}\int d\omega
 \omega^2
\left[if_I f^*_R+h.c.\right],\nonumber\\
f_I&=& \int^\infty_{-\infty} dt e^{2i\omega t} g_I\varphi(t)\nonumber\\
f_R&=&\int^\infty_{-\infty} dt e^{2i\omega t} g_R\varphi(t).
\end{eqnarray}
Here we assume that the oscillation starts at $t=0$ and
 $\varphi(t)$ is given by
\begin{eqnarray}
\varphi(t)&=&\varphi_i e^{-\Gamma t/2} \cos{m t},
\end{eqnarray}
 where $\Gamma$ and $m$ are a decay rate and a mass of $\varphi$ field.
We can calculate the integral, which is given by
\begin{eqnarray}
f_{I(R)}&=&\frac{g_{I(R)}\theta_i}{4i\omega}
\left[
\frac{-\Gamma/2+im} 
{-\Gamma/2+im+2i\omega}\right.\nonumber\\
&&+\left.
\frac{-\Gamma/2-im} 
{-\Gamma/2-im+2i\omega}
\right].
\end{eqnarray}
Since $[if_If_R^*+H.c]=0$ is obvious in this case, the final result
becomes 
\begin{eqnarray}
\Delta n_Q&=&0,
\end{eqnarray}
which suggests that there is no asymmetry generation.
We already know that in the conventional baryogenesis scenario
one has to consider multiple (quantum) corrections to generate the
required interference.
We can see that the same thing is happening in this perturbative
calculation.
(On the other hand, the same interaction can generate the
asymmetry in the nonperturbative limit. 
We will discuss this issue in Sec. \ref{sec-majorana-simpleex} for the
Majorana fermions.)

\subsection{Higher terms for the perturbative expansion}
\label{sec-ql-higher}
To avoid the cancellation, or to introduce interference
between multiple contributions, one can introduce higher terms.
For instance, one can introduce
\begin{eqnarray}
g\Phi(t)&\rightarrow&\Lambda + g_1\varphi + g_2\frac{\varphi^2}{M_*},
\end{eqnarray}
where both $g_1$ and $g_2$ are complex.
Note that this is no longer giving the approximation of the rotational
oscillation.
In the simplest case, $m_R$ can be written as
\begin{eqnarray}
g\Phi(t)&\rightarrow&\Lambda + i \lambda_1 \varphi - \frac{\lambda_2}{2}\frac{\varphi^2}{M_*},
\end{eqnarray}
where $\lambda_i$ is a real constant.
Following the calculation in Ref.\cite{Dolgov:1996qq}, we find
the asymmetry given by
\begin{eqnarray} 
n_\nu-\bar{n}_\nu &=& \frac{1}{16\pi}m
 \Lambda^2\left(\frac{\lambda_1\lambda_2\varphi_i^3}{\Lambda^2M_*}\right)\nonumber\\
&=&\frac{\lambda_1\lambda_2}{16\pi}m
 M_*^2\left(\frac{\varphi_i}{M_*}\right)^3.
\end{eqnarray}
Although the above calculations are useful for understanding the origin
of the asymmetry, the model is a trivial extension of
Ref.\cite{Dolgov:1996qq}.
The only difference is that the terms are not approximating the rotation.

In the followings, we will consider the Dirac and the Majorana fermions
and examine the origin of the asymmetry in the nonperturbative particle
production.

\subsection{The Dirac mass for the nonperturbative calculation}
Usually, the Dirac mass is defined to be real since the redefinition of the
field can remove the phase.
However, if the Dirac mass is time dependent, the Berry connection
appears.
Let us introduce the complex Dirac mass, which is rotating with $m_D(t)\equiv
M_D e^{i\theta(t)}$.
The phase can be removed by defining the Berry
transformation for the left and the right-handed fermions.
In the equation of motion, $\dot{\theta}\ne 0$ introduces asymmetry of
the helicity for each (matter and antimatter)
state, but we will show that there is no asymmetry 
in the total number densities. 
Since the asymmetry is due to the violation of the time-reversal
symmetry by the background, our expectation is that the asymmetry is due to the
shifts of the  ``events'' of the particle production.

To show that our expectation is correct, we start with a simple example.
Since the basic idea of the fermionic preheating has already been discussed in
Refs.\cite{Dolgov:1989us,Greene:1998nh,Chung:1999ve,Peloso:2000hy}, we are going to
follow the notations of Ref.\cite{Peloso:2000hy}.
The new ingredient of our calculation is the complex Dirac mass
\begin{eqnarray}
m_D(t)&=& i\Lambda +g\varphi(t),
\end{eqnarray}
where $g$ is real.
Note that we are not considering a simple circular rotation but a
straight motion, whose orbit is (slightly) shifted from the origin and
introduces significant $\dot{\theta}\ne 0$ when it passes 
near the origin.
Considering the decomposition 
\begin{eqnarray}
\psi&=&\int\frac{d^3k}{(2\pi)^3}
e^{-i\boldsymbol k\cdot \boldsymbol x}\sum_s\left[
u^s_{\boldsymbol k}(t)a^s_{\boldsymbol k}
+v^s_{\boldsymbol k}(t)b^{s\dagger}_{-\boldsymbol k}\right]
\end{eqnarray}
for the Dirac equation
\begin{eqnarray}
(i\slashed{\partial}-m_D)\psi&=&0,
\end{eqnarray}
one will find
\begin{eqnarray}
\dot{u}_\pm&=&ik u_{\mp}\mp i m_D u_\pm,
\end{eqnarray}
which can be decoupled into
\begin{eqnarray}
\label{eq-dec}
\ddot{u}_\pm+\left[\omega^2\pm i\dot{m}_D\right]u_\pm,
\end{eqnarray}
where $\omega(t)^2=k^2+|m_D|^2$.
Let us consider the evolution equation near the bottom,
$\varphi(t_*)=0$.
If we write
\begin{eqnarray}
\varphi(t)&\simeq& \dot{\varphi}_*(t-t_*),
\end{eqnarray}
where $\dot{\varphi}_*$ is a constant defined at $t=t_*$, the equation of
motion gives 
\begin{eqnarray}
\ddot{u}_\pm+\left[k^2+ |i\Lambda +g\dot{\varphi}_*(t-t_*)|^2
\pm ig\dot{\varphi}_*\right]u_\pm=0.
\end{eqnarray}
Obviously, the above Dirac mass $m_D(t)$ does not introduce a new parameter,
which might distinguish $u_\pm$.\footnote{The sign in front
of $ig\dot{\varphi}_*$ does not change the number densities.}
Therefore, the above simplest extension does not introduce asymmetry during
the particle production.
This result reminds us of the perturbative production considered in the
model of $Q$ and $L$. 

\begin{figure}[t]
\centering
\includegraphics[width=1.0\columnwidth]{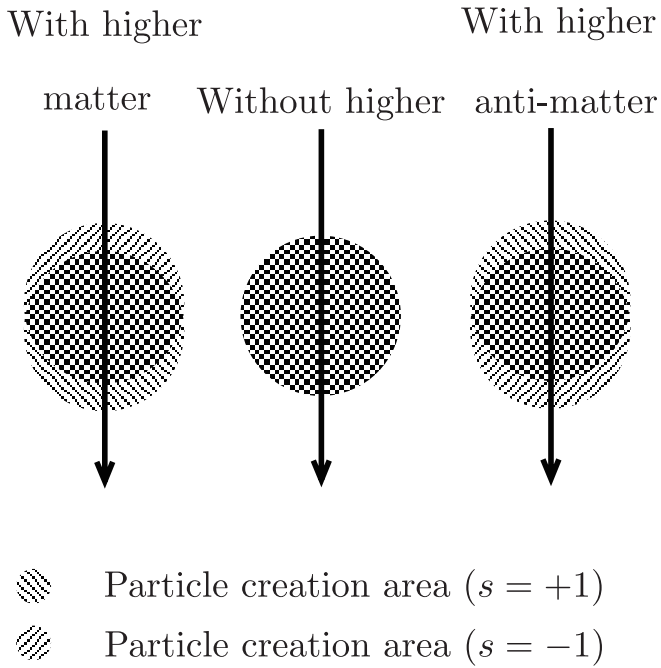}
 \caption{
Particle creation area (nonadiabatic area) for
 $m_D=i\Lambda+g\varphi(t)$ is shown in the middle. 
 In both sides, matter and antimatter creation for
 $m_R=i\Lambda+g_1\varphi+ig_2\frac{\varphi^2}{M_*}$ is shown.
In the left, $s=-1$ creation starts earlier than $s=+1$, while in the
 right, $s=+1$ creation starts earlier than $s=-1$.}\label{fig-dirac}
\end{figure}

To introduce the asymmetry, we consider the higher term, which is given by 
\begin{eqnarray}
m_D(t)&=&i\Lambda + g_1\varphi + ig_2\frac{\varphi^2}{M_*},
\end{eqnarray}
where $\Lambda, g_1,g_2$ are taken to be real.
Then the equation of motion gives 
\begin{eqnarray}
\label{eq-dirac3}
&&\ddot{u}_\pm+\left[k^2+ 
\left|i\Lambda +g_1\dot{\varphi}_*(t-t_*)
+ig_2\frac{\dot{\varphi}^2}{M_*}(t-t_*)^2\right|^2\right.\nonumber\\
&&\left.\pm \left(ig_1\dot{\varphi}_*-2g_2\frac{\dot{\varphi}^2}{M_*}(t-t_*)\right)\right]u_\pm=0.
\end{eqnarray}
Note that the asymmetry appears in the real part in the bracket.
Introducing new parameters 
\begin{eqnarray}
A&\equiv&\dot{\varphi}^2\left(g_1^2+2g_2\frac{\Lambda}{M_*}\right)\nonumber\\
B&\equiv&\dot{\varphi}^2\left(\frac{g_2}{M_*}\right),
\end{eqnarray}
and disregarding $(t-t_*)^4$ near $t=t_*$, one can rewrite the equation as
\begin{eqnarray}
&&\ddot{u}_\pm+\left[k^2+\Lambda^2+ 
A(t-t_*)^2\mp 2B(t-t_*)\pm ig_1\dot{\varphi}_*\right]u_\pm=0.\nonumber\\
\end{eqnarray}
Defining $t_\pm\equiv t_* \pm B/A$ and
\begin{eqnarray} 
\omega_\pm^2&\equiv& k^2+\Lambda^2-B^2/A+
A(t-t_\pm)^2>0,
\end{eqnarray}
we have
\begin{eqnarray}
\ddot{u}_\pm+\left[\omega_\pm^2\pm ig\dot{\varphi}_*\right]u_\pm=0.
\end{eqnarray}
Now the calculation of the number densities is straightforward; the
above equation is almost identical to the standard
equation\cite{Peloso:2000hy}, except for the helicity-dependent $\omega_+\ne
\omega_-$.
The split of $t_*$ into $t_\pm$ ($t_+ >t_*> t_-$) means that the production of
$u_-$ begins earlier than $u_+$.
This is due to the modification of the real part in the bracket in Eq.(\ref{eq-dirac3}). 
Although the nonadiabatic areas are partially overlapping, as is shown
in Fig.\ref{fig-dirac}, it is possible to
expect that the (earlier) production of $u_-$ is so significant that 
it reduces $\dot{\varphi}$ before the (later) production of $u_+$.
In this case, one has to define $\varphi(t)_\pm\simeq
\dot{\varphi}_{\pm}(t-t_\pm)$ for each $u_\pm$.
In the most significant case, where one can assume that almost all the
states in the Fermi sphere are occupied, 
$(|\dot{\varphi}_+|<|\dot{\varphi}_-|)$ determines the
asymmetry of the maximum $|\boldsymbol k|$ of each Fermi sphere.

For the antimatter state, the decoupled equation(\ref{eq-dec}) has
$\dot{m}_D^*$ instead of $\dot{m}_D$. 
Therefore, we find $t_\pm\equiv t_* \mp B/A$ for the antimatter, which
is opposite to the matter, and gives $n_\pm=\bar{n}_\mp$.
In total, there is no asymmetry because $n_++\bar{n}_+=n_-+\bar{n}_-$ is
always satisfied, 
even though $n_+\ne n_-$ and $\bar{n}_+\ne \bar{n}_-$ are possible in this case.

To conclude the particle production due to the Dirac mass, there is no
total asymmetry even if the higher terms are introduced.
The asymmetry of the helicity appears for each (matter and
antimatter) state because the event of the particle production splits.
Similarly, the matter-antimatter asymmetry appears for each helicity
state.
However, these partial asymmetries do not cause generation of the total asymmetry.

To avoid the cancellation of the asymmetries, which has been seen for
the Dirac mass, we will consider the Majorana fermion in the followings.
Note that unlike the Dirac fermion, decoupling of the equations is not
well-defined at the massless point.
First, we consider the rotational oscillation and compare the
perturbative and the non-perturbative particle production.
Then we examine the nonrotational motion.
We will show that unlike the previous models the asymmetry can be generated without
introducing the higher terms. 

\subsection{Majorana fermion for the rotational oscillation}
\label{sec-majorana-osc}
In this section, we consider simple oscillation of $\theta$ for the
Majorana fermion mass.
Using $\Psi^t_R\equiv(\psi_R,\psi_L^c)$, one can write the Majorana mass
term as
\begin{eqnarray}
{\cal L}_m&=&\bar{\Psi}_R
\left(
\begin{array}{cc}
   0   & m_R \\
   m_R^*   & 0 
\end{array}
\right)\Psi_R,
\end{eqnarray}
where we consider $m_R(t)=M_R e^{i\theta(t)}$.
Remember that in Sec. \ref{sec-ql-higher}, we have calculated the asymmetry using the
expansion $e^{i\theta}=1+i\theta +...$.
However, the problem is that the expansion is obviously violating the original
symmetry of the rotation.
To solve this problem, we calculated the asymmetry generation using the
Berry transformation.
The calculational details are shown in Appendix \ref{app-OPE}.
The calculation uses the Yang-Feldman formalism and the Berry transformation.
Thanks to the Berry transformation, we do not have to use the expansion
$e^{i\theta(t)}=[1+i\theta-\theta^2/2+...]$.
Our result shows that the asymmetry appears from the third order of the
perturbative expansion.
Taking the limit $m_\theta \gg |m_\xi| \gg \Gamma$, our result gives the
previous calculation, which is given by 
\begin{eqnarray}
n_\nu-\bar{n}_\nu&\sim&\frac{1}{4\pi}|m_R|^2m_\theta\theta_i^3, \label{eq:net_number_majorana}
\end{eqnarray}
 where $\theta_i$ is an initial phase of the mass.
This result is similar to (\ref{eq:delta_n_q}) if one regards $|m_R|$ as $gf$,
which is a mixing mass term between $Q$ and $L$ in
(\ref{eq:rotated_Lagrangian}).
Note that our calculation takes into the poles shifted by $\Gamma$.

To understand more about the sources of the asymmetry, we will consider
the non-perturbative effect (tunneling) using another schematic
calculation.
We use the Lagrangian given by
\begin{eqnarray}
\label{eq-majorana-Lag}
{\cal L}&=&
\bar{\psi}_Ri\bar{\sigma}^\mu\partial_\mu\psi_R+
\left(m_R\bar{\psi_L^c}\psi_R+m_R^*\bar{\psi_R}\psi_L^c\right).
\end{eqnarray}
The equation of motion is
\begin{eqnarray}
(i\partial_t-i\boldsymbol \sigma \cdot \boldsymbol \partial)\psi_R&=&
-m_R^*\psi_L^c.
\end{eqnarray}
We expand
\begin{eqnarray}
(\psi_R)_\alpha&=&\int\frac{d^3k}{(2\pi)^3}e^{i\mathbf{k\cdot x}}\sum_{s=\pm}
(e^s_{\boldsymbol k})_\alpha \nonumber\\
&&\times \left[
u^s_k(x^0) a^s_{\boldsymbol k}
+s v^s_{\boldsymbol k}(x^0)a_{-\boldsymbol k}^{s\dagger}\right],
\end{eqnarray}
where $e_{\mathbf{k}}^s$ is the eigenstate of the helicity operator, which gives
\begin{equation}
 -k^i \bar{\sigma}^i e_{\mathbf{k}}^s = s|\mathbf{k}| \bar{\sigma}^0 e_{\mathbf{k}}^s.
  \qquad (s=\pm) 
\end{equation}
Substituting this expansion into the equation of motion, we find
\begin{eqnarray}
\label{eq-EOMofMajo}
(i\partial_t+s|\boldsymbol k|)u^s_{\boldsymbol k}=s m_R^*v^{s}_{\boldsymbol k},\nonumber\\
(i\partial_t+s|\boldsymbol k|)v^{s*}_{\boldsymbol k}=-s m_R^* u^{s*}_{\boldsymbol k}.
\end{eqnarray}
Unlike the Dirac fermions, the coefficients of the mixing terms
 (in the right-hand side) are depending on time.
Therefore, to decouple the equations, one has to remove the
time dependence.
After removing the time dependence using the Berry transformation, one
will find the chemical potential and the constant Majorana mass.
The equations can be solved by using the conventional
 decoupling.\footnote{One can find similar calculation in
 Ref.\cite{Adshead:2015kza}, where the decoupling has been used.
 The crucial difference is in the origin of the chemical potential. 
Ref.\cite{Adshead:2015kza} uses the derivative coupling for the chemical
potential, while in the present model it comes from the Berry
connection.
}
Here, we are {\bf not} going to decouple the equations.

This equation can be written as
\begin{eqnarray}
i\frac{d}{dt}\Psi&=&H\Psi,\nonumber\\
\left(
\begin{array}{cc}
H_{11} & H_{12}\\
H_{21} & H_{21}
\end{array}
\right)&=& 
\left(
\begin{array}{cc}
-s|\boldsymbol k| & s m_R^*(t)\\
s m_R(t) & s|\boldsymbol k|
\end{array}
\right),
\end{eqnarray}
where $\Psi^t\equiv (v^s_{\boldsymbol k}, u^s_{\boldsymbol k})$.
For the simple rotational oscillation, we consider 
\begin{eqnarray}
m_R(t)=M_Re^{i\theta(t)},&&
\theta(t)=\theta_0 \cos m_\theta t.
\end{eqnarray}
After the Berry transformation, we find
\begin{eqnarray}
i\frac{d}{dt}\psi^R&=&\hat{H}^R\psi^R,\nonumber\\
\hat{H}^R &=& 
\left(
\begin{array}{cc}
-s|\boldsymbol k| +\frac{\theta_0 m_\theta}{2} \sin m_\theta t& s M_R\\
s M_R & s|\boldsymbol k|-\frac{\theta_0 m_\theta}{2}\sin m_\theta t
\end{array}
\right).\nonumber\\
\end{eqnarray}
The above equation reminds us of the Landau-Zener tunneling\cite{Zener:1932ws}.
\begin{figure}[t]
\centering
\includegraphics[width=1.0\columnwidth]{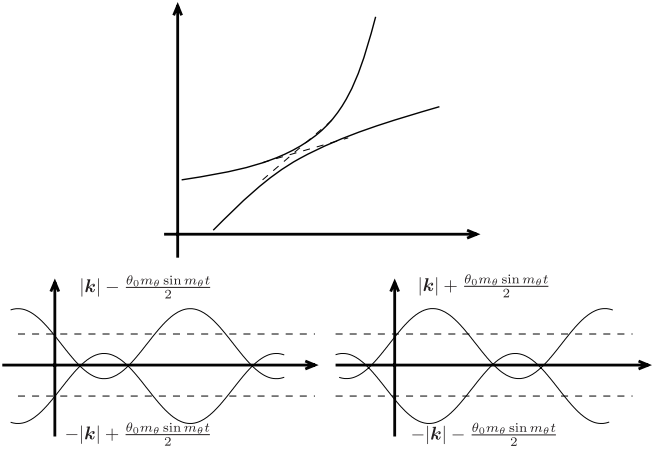}
 \caption{The top figure shows the original Landau-Zener tunneling. 
Eq.(\ref{eq-zener2}) for $s=\pm$ is shown in the bottom.
There is the shift of the  oscillation due to the sign of the helicity.
The tunneling occurs when 
$-s|\boldsymbol k|+\frac{\theta_0^{s(i)} m_\theta}{2}\sin m_\theta
 t_{\boldsymbol k}^{s(i)}=0$. 
From the picture, we can find 
 the relation $t_{\boldsymbol k}^{+(i)}\simeq t_{\boldsymbol
 k}^{-(i)}+\frac{\pi}{m_\theta}$, which is exact when the amplitude does
 not change with time.}\label{fig-zener2}
\end{figure}
Following Ref.\cite{Zener:1932ws}, we can define
\begin{eqnarray}
\label{eq-zener2}
&&\left(
\begin{array}{cc}
\epsilon_1 & \epsilon_{12}\\
\epsilon_{21} & \epsilon_{2}
\end{array}
\right)\nonumber\\
&=& 
\left(
\begin{array}{cc}
-s|\boldsymbol k|  +\frac{\theta_0 m_\theta}{2}\sin m_\theta t& s M_R\\
s M_R & s|\boldsymbol k| -\frac{\theta_0 m_\theta}{2}\sin m_\theta t
\end{array}
\right),
\end{eqnarray}
which is shown in Fig.\ref{fig-zener2} together with the original Landau-Zener
tunneling.
The probability of the translation at each
crossing point is given by
\begin{eqnarray}
\label{eq-zener-p}
{\cal P}_{\boldsymbol k}^{s(i)}&\simeq&
 e^{-\pi p^{s(i)}_{\boldsymbol k}},\nonumber\\
p_{\boldsymbol
 k}^{s(i)}&\equiv&\frac{M_R^2}{\left|\frac{\theta_0^{s(i)}
				m_\theta^2}{2}\cos m_\theta t_{\boldsymbol k}^{s(i)}\right|},
\end{eqnarray}
where $t_{\boldsymbol k}^{s(i)}$ for the ith event is defined by
\begin{eqnarray}
\label{eq-i-th-time}
 -s|\boldsymbol k|  +\frac{\theta_0^{s(i)} m_\theta}{2}\sin
  m_\theta t_{\boldsymbol k}^{s(i)}&=&0.
\end{eqnarray}
Substituting Eq.(\ref{eq-i-th-time}) into
Eq.(\ref{eq-zener-p}), we find 
\begin{eqnarray}
\label{eq-p-zener}
p_{\boldsymbol
 k}^{s(i)}&\equiv&\frac{2M_R^2}{m_\theta^2\theta_0^{s(i)}}
 \left(1-\frac{4|\boldsymbol k|^2}{m_\theta^2 (\theta_0^{s(i)})^2}
\right)^{-1/2},
\end{eqnarray}
where $|\boldsymbol k|< \theta_0^{s(i)}m_\theta/2$ is required for the tunneling.
Using the conventional Bogoliubov transformation, the total number density
can be calculated as 
\begin{eqnarray}
n^{s}_{\boldsymbol k}&\simeq&\sum_i{\cal P}_{\boldsymbol k}^{s(i)},
\end{eqnarray}
where particles are assumed to decay before the next particle production.
One can verify that the above result is consistent with Refs.\cite{Peloso:2000hy,
Adshead:2015kza}, in which the Landau-Zener tunneling has not
been used for the calculation. 
Since $m_\theta$ is the mass of $\theta$ appearing in
$m_R=M_Re^{i\theta}$, the limit $M_R\ll m_\theta$ is unlikely in this
model. 
Note also that the equation becomes singular in the limit
$M_R\rightarrow 0$ (See Ref.\cite{Zener:1932ws}).

Let us temporarily assume that the amplitude $\theta_0$ is constant
within a cycle.
Then, one can see that the same particle
creation is occurring for each helicity state, but they do
not happen simultaneously.
From Fig.\ref{fig-zener2}, one can see that the particle creation is
delayed for the helicity state $s=-1$, and  
the delay is just a half of the oscillation time.

When the amplitude decreases with time, the delay is approximately given
by $\Delta t\simeq \frac{\pi}{m_\theta}$.
Then, if the amplitude behaves like $\propto e^{-\Gamma t}$,
the amplitude $\theta_0^{s(i)}$ defined for the ith event can be
expressed as
\begin{eqnarray}
\theta_0^{-(i)}&\simeq&e^{-\Gamma \Delta t}\theta_0^{+(i)}.
\end{eqnarray}
In this case, the origin of the asymmetry is $\theta_0^{-(i)}\ne
\theta_0^{+(i)}$, which directly biases $p_{\boldsymbol k}^{s(i)}$ in Eq.(\ref{eq-p-zener}).
Therefore, we can clearly understand that the time-dependent amplitude
is the source of the asymmetry in this case.

From Eq.(\ref{eq-p-zener}), one can see that $p_{\boldsymbol
k}^{s(i)}$ can be approximated as a constant within
$|\boldsymbol k|< k_{Max}^{(i)}\equiv\theta_0^{s(i)}m_\theta/2$.
If one assumes $p_{\boldsymbol k}^{s(i)}\sim 1$ within 
$|\boldsymbol k|< k_{Max}$, the asymmetry becomes
\begin{eqnarray}
n_+^{(i)}-n^{(i)}_-&\propto& (k^{(i)}_{Max})^2\frac{dk^{(i)}_{Max}}{d \theta_0^+}d\theta_0^+\nonumber\\
&\sim&m_\theta^2 \Gamma (\theta_0^{+(i)})^3,
\end{eqnarray}
where we expanded 
\begin{eqnarray}
\theta_0^{-(i)}&\simeq&\left(1-\frac{\pi\Gamma}{m_\theta}\right)\theta_0^{+(i)}.
\end{eqnarray}
Here, we assumed that $\Gamma\propto m_\theta$ and $\Gamma \Delta
t\simeq \frac{\pi\Gamma}{m_\theta}\ll 1$.

\subsection{Majorana Fermions for the simplest extension}
\label{sec-majorana-simpleex}
Instead of considering the rotational motion, we are going to 
introduce the Majorana mass given by
\begin{eqnarray}
m_R(t)&=& i\Lambda + g \varphi(t),
\end{eqnarray}
where (just for simplicity) both $\Lambda$ and $g$ are taken to be real.
Previously, we have seen that the above extension (without higher
terms) does not generate the asymmetry for the perturbative calculation.

In this section, using the non-perturbative calculation, 
we will show that the above extension can generate the asymmetry.
We consider significant particle
production, which is realized when the oscillation starts with 
$|\Lambda/g\varphi|\ll 1$.
Again, we use the Lagrangian given by Eq.(\ref{eq-majorana-Lag}).
To decouple the equation of motion (\ref{eq-EOMofMajo}) for
$\dot{m}_R\ne 0$, we rewrite the equation as\footnote{This manipulation
is {\bf not possible in the standard calculation of preheating}, since
$m_R$ is usually assumed to be a real parameter and the particle
production is considered around $m_R\simeq 0$.
Ref.\cite{Pearce:2015nga} considers a model with a time-dependent
chemical potential (from the derivative coupling) and a time-dependent
(real) Majorana mass.
Their first equations coincide with our equations after using the
Berry transformation.
However, because of the (possible) appearance of $m_R=0$,
 their secondary equations do not coincide with our calculation.} 
\begin{eqnarray}
(m_R^*)^{-1}(i\partial_t+s|\boldsymbol k|)u^s_{\boldsymbol k}=s v^{s}_{\boldsymbol k},
\end{eqnarray}
and obtain
\begin{eqnarray}
&&(-i\partial_t+s|\boldsymbol
 k|)\left[(m_R^*)^{-1}(i\partial_t+s|\boldsymbol k|)u^s_{\boldsymbol
     k}\right]\nonumber\\
&=&\left[\frac{\partial_t^2+|\boldsymbol k|^2}{m_R^*}
+i\frac{\dot{m}_R^*}{m_R^{*2}}
(i\partial_t+s|\boldsymbol k|)\right]
u^s_{\boldsymbol k}\nonumber\\
&=&-m_R u^{s}_{\boldsymbol k}.
\end{eqnarray}
Therefore, one obtains the decoupled equation 
\begin{eqnarray}
\left[\partial_t^2
-\frac{\dot{m}_R^*}{m_R^*}\partial_t 
+|\boldsymbol k|^2  +i s|\boldsymbol k|\frac{\dot{m}_R^*}{m_R^*}
\right]
u^s_{\boldsymbol k}
&=&-|m_R|^2 u^{s}_{\boldsymbol k}.\nonumber\\
\end{eqnarray}
Substituting
\begin{eqnarray}
u_{\boldsymbol k}^s&=&e^{\int \frac{\dot{m}_R^*}{2m_R^*} dt}
 U_{\boldsymbol k}^s,
\end{eqnarray}
we find
\begin{eqnarray}
&\ddot{U}_{\boldsymbol k}^s+\left[
\dot{\left(\frac{\dot{m}_R^*}{m_R^*}\right)}-\frac{1}{4}\left(\frac{\dot{m}_R^*}{m_R^*}\right)^2
+|\boldsymbol k|^2+|m_R|^2+is|\boldsymbol k|\frac{\dot{m}_R^*}{m_R^*}
\right]U&\nonumber\\
&=0&.
\end{eqnarray}
From this equation, one can immediately understand that the asymmetric
particle production is possible in this case.
If we define $v\equiv g\dot{\phi}(t_*)$, 
$g\varphi(t)=v(t-t_*)$ near the bottom of the oscillation, we have 
\begin{eqnarray}
\frac{\dot{m}_R^*}{m_R^*}\simeq \frac{v}{i\Lambda},&&
\dot{\left(\frac{\dot{m}_R^*}{m_R^*}\right)}
\simeq \frac{v^2}{\Lambda^2}
\end{eqnarray}
Then, for $t_*=0$, we can write the equation as
\begin{eqnarray}
\ddot{U}_{\boldsymbol k}^s+\left[
\frac{5}{4}\frac{v^2}{\Lambda^2}
+|\boldsymbol k|^2+\Lambda^2+v^2 t^2
+s|\boldsymbol k|\left(\frac{v}{\Lambda}\right)
\right]U&=&0.\nonumber\\
\end{eqnarray}
Defining 
\begin{eqnarray}
|\boldsymbol \hat{k}^s|&\equiv& |\boldsymbol k|
 +\frac{s}{2}\frac{v}{\Lambda},
\end{eqnarray}
we obtain 
\begin{eqnarray}
\ddot{U}_{\boldsymbol k}^s+\left[
|\boldsymbol \hat{k}^s|^2+m_{eff}^2(t)
\right]
U&=&0,
\end{eqnarray}
where 
\begin{eqnarray}
m_{eff}^2(t)\equiv
\frac{v^2}{\Lambda^2}+\Lambda^2+v^2 t^2. 
\end{eqnarray}
Therefore, when $\Lambda^2\ll v$, particle production is not significant
near the bottom.
The simplest assumption, which justifies the significant particle
production, is $\Lambda\sim v^{1/2}$. 
The equation is almost the same as the conventional preheating,
except for the helicity-dependent $|\boldsymbol \hat{k}^s|^2$.
We can conclude that the asymmetry is due to the split of
$|\boldsymbol \hat{k}^s|^2$.
During the particle production, it shifts the radius of the Fermi
sphere of each helicity state. 

Unlike the perturbative expansion discussed in
Sec. \ref{sec-small-pert}, our non-perturbative calculation gives 
the asymmetry.
Basically, the perturbative expansions and the non-perturbative effects
(such as the tunnelings) will give different contributions.
These are expected to be unified in
the resurgence theory\cite{Delabaere:1999,
Dorigoni:2014hea}.
Since the basic equations are written in ordinary differential equations (ODEs),
one can use the resurgence of ODEs, which has already been solved in the
mathematical side. 
The task is to identify the origin of the asymmetry in the
framework of the resurgence.
Note that $i\Lambda \rightarrow -i\Lambda$ flips the asymmetry and there
is a singularity at $\Lambda=0$.
The relation will be revealed in our next paper.

\subsection{Comment on a more ambitious approach: Multifield extension and the Cabibbo-Kobayashi-Maskawa matrix}
\label{sec:morefields}
In the above models, the source of the phase is designed to be 
very simple.
The phase in the off-diagonal element determines the Berry
phase, and there is the obvious correspondence between them.
We have also seen that a simple extension of the scenario (i.e,
$\Delta=\Lambda+ig\varphi$) can be used to generate the effective
chemical potential.
In this case, there is no obvious correspondence between the Berry phase
and the phases of the ``fundamental'' parameters $\Lambda, g$, and $\varphi$.
However, generation of the effective chemical potential is
very clear in the light of the Berry transformation.

In the above models, all phases in the
Hamiltonian can be removed by the field rotation, which we called ``the
Berry transformation.''
Now our question is very simple.
``What happens if the fields are multiplied and the Berry transformation
has to be given by a complex function of the original parameters?''

One can examine the above idea in the three-family fermion model.
One can introduce the flavor index $i=1,2,3$ and write
\begin{eqnarray}
{\cal L}&=& i\bar{Q}_{i}(i\gamma^\mu\partial_\mu-m_{Q}^{ij})Q_j
+\bar{L}_i(i\gamma^\mu\partial_\mu-m_{L}^{ij})L_j\nonumber\\
&&+\left(f^{ij}\bar{Q}_i L_j + H.c.\right),
\end{eqnarray}
where $m_Q, m_L$ can be diagonalized by unitary transformations
where 
$\hat{Q}_i = (U_Q^\dagger)^{ij}Q_j, \hat{L}_i=(U_L^\dagger)^{ij}L_j$.
Then the interaction is written as
\begin{eqnarray}
{\cal L}_{int}&=& \left( (U_Q^\dagger f U_L)^{ij}\bar{\hat{Q}}_i\hat{L}_j + H.c.\right).
\end{eqnarray}

Now the CP phase appearing in the matrix $V^{ij}\equiv (U_Q^\dagger f U_L)^{ij}$ is 
quite similar to the famous Cabibbo-Kobayashi-Maskawa matrix in the
standard model.
Unlike the naive $2\times 2$ matrix models we have considered
previously, the phases in $V^{ij}$ are not simply determined by the
complex phases of the original parameters of the Lagrangian.
(To avoid confusions, note that we have considered only the single
flavor for the Q-L model. 
$2\times 2$ matrix was considered for the kaonlike models and
the Majorana fermions, but the matrix was for the matter and
the antimatter, not for the flavor.)
The phases in $V^{ij}$ are given by the functions of all the original parameters.
Therefore, even if the time-dependent motion 
does not accompany any rotation of the original complex parameter,  
the motion may introduce time-dependent phases in $V^{ij}$, which
can eventually introduce the chemical potential in the Hamiltonian
through the Berry connection.
Note that usually the phases in $V^{ij}$ are removed by the field
rotations and only a CP phase (Kobayashi-Maskawa CP phase) remains.

The minimal multifield extension that realizes the above idea is given
by a complex scalar field couples to a real scalar field.
Consider the following Lagrangian;
\begin{eqnarray}
{\cal L}&=&|\partial_\mu \phi|^2 -m_\phi^2|\phi|^2\nonumber\\
&& +\frac{1}{2}(\partial_\mu\eta)^2-\frac{1}{2}m_\eta^2
 \eta^2\nonumber\\
&&-\frac{1}{2}(\epsilon\phi^2+H.c.) -(g\phi+H.c)\eta,
\end{eqnarray}
where $\phi$ is a complex scalar and $\eta$ is a real scalar.
Here, $\epsilon$, $g$ are complex coupling constants.
Note that in this case the complex phases of $\epsilon$ and $g$ are not
removed simultaneously by the field rotation.
Therefore, at least one complex phase will remain ``after'' the field rotation.
The equations of motion are given by
\begin{equation}
 \ddot{\Psi}^R+\Omega^2\Psi^R=0 \label{eq:eom_phichi}
\end{equation}
where
\begin{equation}
 \Psi^R \equiv \left( \begin{array}{cc} \phi \\ \phi^\dagger \\ \eta \end{array} \right),
 \quad \Omega^2 \equiv \left(
  \begin{array}{ccc}
   \omega_\phi^2 & \epsilon_R & g^*\\
   \epsilon_R & \omega_\phi^2 & g\\
   g & g^* & \omega_\eta^2
  \end{array}
 \right).
\end{equation}
Here, $\epsilon_R$ is set real but $g$ is still a complex parameter.
Note that we have already used the field rotation of $\phi$ to have
$\epsilon_R$. 
To diagonalize the matrix using a unitary matrix $U$, one has to
calculate eigenvectors of the matrix $\Omega^2$.
Because $\Omega^2$ has a complex element, the unitary matrix must also
have complex elements.
Here, the key idea is that the unitary matrix can be decomposed using
(real) rotation and complex matrices\cite{Kobayashi:1973fv}, which are
sometimes denoted as $U_{12}, U_{13}, U_{23}$ and $U_{\theta_i}$. 
Since the particle production occurs for the eigenstates, one has to
consider $\Psi^E$, which is given by the transformation using 
 $U_{12}, U_{13}, U_{23}$ and $U_{\theta_i}$.
In Ref.\cite{Enomoto:2017rvc}, we have shown that the ``eigenstates''
are preserving matter-antimatter asymmetry but they are mixed by the
Berry connection.
In this case, the phases in the Berry transformation are functions of 
$\omega_\phi, \epsilon_R, \omega_\eta$ and $g$.
Therefore, in this model, one can expect that a time-dependent
$\omega_\phi$ can generate matter-antimatter asymmetry,
 since it may change the phase parameter as
 $\dot{\theta}_i=\dot{\omega}_\phi(\partial\theta_i/\partial
 \omega_\phi)$. 
Note that $\omega_\phi$ itself does not have a phase, which is similar
to the simple extension discussed in Sec. \ref{sec:exten}.
The analytic relation between the chemical potential and
$\dot{\omega}_\phi$ has a very lengthy form, since it uses the
eigenvectors of the $3\times 3$ matrix $\Omega^2$.
In Ref.\cite{Enomoto:2017rvc}, we showed a numerical calculation to show
that the matter-antimatter asymmetry is generated in this model.
Viewing with the ``eigenstates''($\Psi^E$), the Berry connection causes
mixing between ``eigenstates'' accompanied by the CP phase, which is
time dependent,  to generate the interference between states.

\section{The Berry phase and the Legendre transformation}
\label{sec-legendre}
The chemical potential may cause a problem in the Legendre transformation if it is
explained by the derivative coupling of a field in motion.
The reason is very simple.
If the chemical potential is introduced using the derivative of the
field $\phi$,
the Lagrangian density acquires the term
\begin{eqnarray}
{\cal L}_c&=& (\partial_\mu \phi)J^\mu_\phi,
\end{eqnarray}
which shifts the conjugate momentum given by
\begin{eqnarray}
\pi&=&\frac{\partial {\cal L}}{\partial \dot{\phi}}.
\end{eqnarray}
Since the corresponding part of the Hamiltonian density is 
\begin{eqnarray}
{\cal H}_c&=&\pi \dot{\phi}-{\cal L}_c,
\end{eqnarray}
the chemical potential disappears from the Hamiltonian.
The problem has been discussed in Refs.\cite{Arbuzova:2016qfh, Dasgupta:2018eha}.

In this section, we are going to show a more transparent consistency
relation between the Berry connection and the Legendre transformation.
It is easy to see that the chemical potential defined using the Berry
connection appears in the Hamiltonian (after the Legendre transformation)
in the expected form.
Note also that the Berry transformation and the Legendre transformation
obviously commute. 
We start with the Lagrangian\cite{Arbuzova:2016qfh}:
\begin{eqnarray}
{\cal L}&=&g^{\mu\nu}\partial_\mu\Phi^*\partial_\nu\Phi-V(\Phi^*\Phi)+
\bar{Q}(i\gamma^\mu\partial_\mu-m_Q)Q\nonumber\\
&&+\bar{L}(i\gamma^\mu\partial_\mu-m_L)L+
{\cal L}_{int}(\Phi,Q,L)\nonumber\\
{\cal L}_{int}&=&\frac{\sqrt{2}}{m_X^2}\frac{\Phi}{f}
\left(\bar{L}\gamma_\mu Q\right)\left(\bar{Q}^c\gamma_\mu Q\right)+H.c.
\end{eqnarray}

Let us consider the Berry transformation.
We define
\begin{eqnarray}
Q_E\equiv U^{-1}Q&,
\end{eqnarray}
where the Berry transformation is defined by
$U(\alpha)=e^{-i\alpha/3}$ with an arbitrary parameter $\alpha$.
Inserting $UU^{-1}=1$ in front of $Q$, one will find
\begin{eqnarray}
{\cal
 L}&=&g^{\mu\nu}\partial_\mu\Phi^*\partial_\nu\Phi-V(\Phi^*\Phi)
\nonumber\\
&&+\bar{Q}_E(i\gamma^\mu\partial_\mu-m_Q)Q_E+\bar{L}(i\gamma^\mu\partial_\mu-m_L)L
\nonumber\\
&&+{\cal L}_{int}(\Phi,Q_E,L)+{\cal L}_{chem}\nonumber\\
{\cal L}_{int}&=&\frac{\sqrt{2}}{m_X^2}\frac{\Phi e^{-i\alpha}}{f}
\left(\bar{L}\gamma_\mu Q_E\right)\left(\bar{Q}_E^c\gamma_\mu Q_E\right)+H.c,\nonumber\\
{\cal L}_{chem}&=&(\partial_\mu \alpha)J^\mu.
\end{eqnarray}
where $J_\mu =(1/3)\bar{Q}_E\gamma_\mu Q_E$ is the baryon current.
Consider the classical rotational motion of the field $\Phi\equiv fe^{i\theta}$.
On the orbit of the rotational motion, the phase of $\Phi e^{-i\alpha}=
fe^{i(\theta-\alpha)}$ in ${\cal L}_{int}$
can be fixed by choosing the arbitrary parameter $\alpha$.
When the parameter $\alpha$ is chosen to make the phase constant on
the orbit, $\alpha$ has to be changing along the orbit.
In this case, one can see that the classical rotational
motion of the field $\Phi$ introduces the effective chemical potential,
which is nothing but the Berry connection.

Using the Legendre transformation, one can calculate the
Hamiltonian of the system.
Since the Berry transformation is nothing but inserting ``$1=UU^{-1}$'' in
front of the field, as we have explained above, the Legendre
transformation and the Berry transformation must commute.
In the above example, it is obvious that the same chemical
potential appears in the Hamiltonian. 
On the other hand, if one identifies the ``parameter'' $\alpha$ with
the Nambu-Goldstone ``field'', these manipulations (Legendre
transformation and the Berry transformation) do not
commute.

As is already discussed in Sec. \ref{sec-decay},
the phase in ${\cal L}_{int}$ can be fixed
by defining the Berry transformations as
\begin{eqnarray}
Q_E\equiv e^{is\alpha/3}Q, && L_E\equiv e^{-i(1-s)\alpha}L
\end{eqnarray}
and adjusting the parameter $\alpha$ to cancel the phase in ${\cal
L}_{int}$.
In light of the Berry phase, there seems to be no obvious
reason that one has to choose a priori the specific value $s=1$.

\section{Conclusions and discussions}

In this paper, we examined the spontaneous baryogenesis scenario using
the framework of the Berry phase.
In this approach, the chemical potential is not the
derivative coupling of the Nambu-Goldstone boson but the Berry
connection defined for the ``Berry transformation''.
In this paper, the ``Berry transformation'' is defined specifically for
the transformation, which removes the phase in the Hamiltonian during
the evolution.

The merit of this approach is the obvious consistency between the
Hamiltonian and the Lagrangian formalisms.
The Berry transformation commutes with the Legendre transformation, and
the chemical potential in the thermal equilibrium is obvious in this
approach.

Then, using the Berry transformation as a useful tool for the calculation, 
we examined the asymmetry generation during the particle production in 
time-dependent backgrounds.
In the framework of the Berry phase, the chemical potential is given by
the Berry connection associated with the conventional Berry phase.
The conventional Berry phase may appear both in the adiabatic and in the
nonadiabatic evolutions.
On the other hand, the particle production in the time-dependent
background is caused by the transition between states.
In the framework of the Berry phase, this can introduce the
nonadiabatic Berry phase, which appears only in the nonadiabatic
evolution and vanishes in the adiabatic limit.
In this paper, we compared the perturbative and the
non-perturbative calculations.
Our speculation is that the asymmetry in the non-perturbative particle
production can be explained by the resurgence
theory\cite{Delabaere:1999, Dorigoni:2014hea}.

Besides the discrepancy between the perturbative and the
non-perturbative calculations, we also examined the effect of the
expansion $e^{i\theta}=[1+i\theta-\theta^2/2+...]$, which is explicitly
violating the original symmetry.
In our calculation, the Berry transformation is very useful since it 
enabled us to calculate the asymmetry without using the above expansion. 

For the rotational oscillation of the time-dependent Majorana mass term,
we calculated the non-perturbative particle production using the Landau-Zener
tunneling. 
In this case, the non-perturbative calculation is explicitly defined for
the tunneling process and the source of the asymmetry is the
split of the tunneling.

The model can also be extended to multifield models, in which the
Berry phases are complex functions of the original parameters.
Although the parameter dependence of the CP phase becomes very
complicated compared with the original spontaneous baryogenesis
scenario, theoretically one can decompose the unitary matrix in the
simpler form to find that $\dot{U}_{\theta_i}\ne 0$ is the source of the
matter-antimatter asymmetry. 

From the results, we found that the Berry phase and the Berry connection
are giving a natural framework of the spontaneous baryogenesis scenario.
The asymmetry of perturbative and non-perturbative particle production
will be understood in the resurgence theory.
Although in this paper we have considered only a time-dependent
parameter, one can also consider a ``space''-dependent parameter as the
source of the Berry connection, which 
may appear on topological defects such as walls and strings.

\section{Acknowledgments}
SE is supported by the Heising-Simons Foundation grant No 2015-109,
and by JSPS KAKENHI Grant Numbers JP18H03708, JP17H01131.

\appendix

\section{Net number for Majorana fermion}
\label{app-OPE}
In this section, we derive the formula (\ref{eq:net_number_majorana})
which is the net number density induced by the varying phase of the mass.
At first we start with the following Lagrangian:
\begin{eqnarray}
 \mathcal{L}
  &=& \frac{1}{2}f^2(\partial\theta)^2-\frac{1}{2}m_\theta^2f^2\theta^2
   +\xi^{\dagger} i \bar{\sigma}^\mu\partial_\mu \xi\nonumber\\
  & & \quad - \frac{1}{2} |m_\xi|e^{i\theta(t)}\xi\xi
   - \frac{1}{2} |m_\xi|e^{-i\theta(t)} \xi^{\dagger} \xi^{\dagger}
\end{eqnarray}
where $\theta=\theta(t)$ is a real $c$-number field, $\xi$ is a two-component spinor field,
and $f$ is a constant.  Next, we remove the phase of the mass by taking $\xi_E\equiv e^{i\theta/2}\xi$.
Then the Lagrangian becomes to
\begin{eqnarray}
 \mathcal{L}
  &=& \frac{1}{2}f^2(\partial\theta)^2-\frac{1}{2}m_\theta^2f^2\theta^2
   +\xi_E^{\dagger} i \bar{\sigma}^\mu\partial_\mu \xi_E\nonumber\\
  & & \quad - \frac{1}{2} |m_\xi|\xi_E\xi_E - \frac{1}{2} |m_\xi| \xi^\dagger_E \xi^\dagger_E
   +\frac{1}{2}\partial_\mu\theta\cdot\xi_E^\dagger\bar{\sigma}^\mu\xi_E. \nonumber \\
\end{eqnarray}
Although the phase in the mass term disappears,
note that there appears the additional term associated with $\partial_\mu\theta$
which corresponds to a part of the chemical potential.
From this Lagrangian, the equations of motion are given by
\begin{eqnarray}
 0 &=& i\bar{\sigma}^\mu\partial_\mu\xi_E-|m|\xi^\dagger_E
  +\frac{1}{2}\partial_\mu\theta\cdot\bar{\sigma}^\mu\xi_E \label{eq:EOM_xi} \\
 0 &=& i\sigma^\mu\partial_\mu\xi^\dagger_E-|m|\xi_E
  -\frac{1}{2}\partial_\mu\theta\cdot\sigma^\mu\xi_E^\dagger \label{eq:EOM_xi_dagger} \\
 0 &=& f^2\partial^2\theta
  +\frac{1}{2}\partial_\mu\left(\xi_E^\dagger \bar{\sigma}^\mu\xi_E\right)+f^2m_\theta^2\theta \label{eq:EOM_theta}
\end{eqnarray}
Instead of using (\ref{eq:EOM_theta}), we use an approximated equation
\begin{equation}
 \ddot{\theta} + \Gamma\dot{\theta}+ m_\theta^2 \theta=0 \label{eq:eom_theta_approx}
\end{equation}
where $\Gamma$ is a decay rate of $\theta$.
(\ref{eq:EOM_xi}) and (\ref{eq:EOM_xi_dagger}) are equivalent.

Using these equations of motion, we will calculate the number density with Yang-Feldman formalism
where the operator field is represented by an asymptotic filed and Green function.

\subsection{Yang-Feldman equation}

The formal solution called as Yang-Feldman equation for Eq.(\ref{eq:EOM_xi}) and (\ref{eq:EOM_xi_dagger})
are given by
\begin{eqnarray}
 \left( \begin{array}{c} \xi_E(x) \\ \xi_E^\dagger(x) \end{array} \right)
  &=& \left( \begin{array}{c} \xi_E^{\rm in}(x) \\ \xi_E^{{\rm in}\dagger}(x) \end{array} \right) \nonumber \\
  & & - \int d^4y \: G_{xy}^{\rm in} \left(
   \begin{array}{c} -\frac{1}{2}\dot{\theta}(y^0)\cdot\bar{\sigma}^0\xi_E(y) \\
    \frac{1}{2}\dot{\theta}(y^0)\cdot\sigma^0\xi_E^\dagger(y) \end{array} \right) \nonumber \\ \label{eq:YF_eq}
\end{eqnarray}
where $\xi_E^{\rm in}$ is an asymptotic field which is defined at $x^0=t^{\rm in}$ and satisfies
\begin{eqnarray}
 0 &=& i\bar{\sigma}^\mu\partial_\mu\xi_E^{\rm in}-|m|\xi^{{\rm in}\dagger}_E, \\
 0 &=& i\sigma^\mu\partial_\mu\xi^{{\rm in}\dagger}_E-|m|\xi_E^{\rm in}.
\end{eqnarray}
Since $\xi_E^{\rm in}$ is same to a free field, we can expand it to
\begin{eqnarray}
 (\xi_E^{\rm in})_\alpha(x)
  &=& \int\frac{d^3k}{(2\pi)^3}e^{i\mathbf{k\cdot x}}\sum_{s=\pm} (e_{\mathbf{k}}^s)_\alpha \tilde{\xi}_{\mathbf{k}}^s(x^0)
   \label{eq:asym_Majorana_fermion} \\
 \tilde{\xi}_{\mathbf{k}}^s(x^0) & \equiv & u_k^s(x^0) a_{\mathbf{k}}^s
   +se^{-i\rho_{\mathbf{k}}} v_k^{s*}(x^0) a_{-\mathbf{k}}^{s\dagger} \label{eq:Majorana_k_rep}
\end{eqnarray}
where $e_{\mathbf{k}}^s$ is an eigenvector for helicity state which is defined by
\begin{equation}
 -k^i \bar{\sigma}^i e_{\mathbf{k}}^s = s|\mathbf{k}| \bar{\sigma}^0 e_{\mathbf{k}}^s.
  \qquad (s=\pm) \label{eq:def_helicity}
\end{equation}
In this paper we choose a representation satisfying (\ref{eq:def_helicity}) as
\begin{eqnarray}
 (e_{\mathbf{k}}^s)_1 &=& \sqrt{\frac{1}{2}\left(1+\frac{sk^3}{|\mathbf{k}|}\right)},\\
 (e_{\mathbf{k}}^s)_2 &=& se^{i\rho_{\mathbf{k}}}\sqrt{\frac{1}{2}\left(1-\frac{sk^3}{|\mathbf{k}|}\right)}, \\
 e^{i\rho_{\mathbf{k}}} & \equiv & \frac{k^1+ik^2}{\sqrt{(k^1)^2+(k^2)^2}} .
\end{eqnarray}
Then the eigenvector satisfies following relations:
\begin{equation}
 e_{\mathbf{k}}^{s\dagger}\bar{\sigma}^0e_{\mathbf{k}}^r=
 e_{\mathbf{k}}^s\sigma^0e_{\mathbf{k}}^{r\dagger}=\delta^{sr},
\end{equation}
\begin{equation}
 (e_{-\mathbf{k}}^{s\dagger})^{\dot{\alpha}}
  =-se^{-i\rho_{\mathbf{k}}}(\bar{\sigma}^0e_{\mathbf{k}}^s)^{\dot{\alpha}},
\end{equation}
\begin{eqnarray}
 (e_{\mathbf{k}}^s)_\alpha(e_{\mathbf{k}}^{s\dagger})_{\dot{\alpha}}
  &=& \frac{1}{2}\left(\sigma^0+\frac{sk^i}{|\mathbf{k}|}\sigma^i\right)_{\alpha\dot{\alpha}}
   \quad (\text{no sum for } s) \nonumber \\ \\
 (e_{\mathbf{k}}^{s\dagger})^{\dot{\alpha}}(e_{\mathbf{k}}^s)^\alpha
  &=& \frac{1}{2}\left(\bar{\sigma}^0+\frac{sk^i}{|\mathbf{k}|}\bar{\sigma}^i\right)^{\dot{\alpha}\alpha}
   \quad (\text{no sum for } s) \nonumber \\
\end{eqnarray}
In Eqs.(\ref{eq:Majorana_k_rep}), $a_{\mathbf{k}}^s$ ($a_{-\mathbf{k}}^{s\dagger}$) is an annihilation (a creation) operator satisfying
\begin{equation}
 \{a_{\mathbf{k}}^s, a_{\mathbf{k}'}^{s'\dagger}\}=(2\pi)^3\delta(\mathbf{k}-\mathbf{k}'),
  \quad (\text{others}) =0,
\end{equation}
and $u_k^s$, $v_k^{s*}$ are time-dependent parts of wave functions whose solutions for zero particle state
is given by
\begin{eqnarray}
 u_k^s(x^0) &=& \sqrt{\frac{1}{2}\left(1-\frac{s|\mathbf{k}|}{\omega_k}\right)}e^{-i\omega_kx^0} \\
 v_k^{s*}(x^0) &=& \sqrt{\frac{1}{2}\left(1+\frac{s|\mathbf{k}|}{\omega_k}\right)}e^{+i\omega_kx^0} \\
 \omega_k &\equiv& \sqrt{|\mathbf{k}|^2+|m_\xi|^2}.
\end{eqnarray}
Moreover, in (\ref{eq:YF_eq}),
\begin{eqnarray}
 G_{xy}^{\rm in}
  &\equiv& i\left(\theta(x^0-y^0)-\theta(t^{\rm in}-y^0)\right) \nonumber \\
  & & \times \left( \begin{array}{cc}
     \{\xi_E^{\rm in}(x),\xi_E^{{\rm in}\dagger}(y)\} & \{\xi_E^{\rm in}(x),\xi_E^{\rm in}(y)\} \\
     \{\xi_E^{{\rm in}\dagger}(x),\xi_E^{{\rm in}\dagger}(y)\} & \{\xi_E^{{\rm in}\dagger}(x),\xi_E^{\rm in}(y)\}
    \end{array}
    \right) \nonumber \\
  &=& i\left(\theta(x^0-y^0)-\theta(t^{\rm in}-y^0)\right) \nonumber \\
  & & \times \int\frac{d^3k}{(2\pi)^3}e^{i\mathbf{k}\cdot(\mathbf{x-y})}\sum_s \nonumber \\
  & & \times \left( \begin{array}{cc}
     e_{\mathbf{k}}^s & \\ & \bar{\sigma}^0e_{\mathbf{k}}^s \end{array} \right) F_k^s(x^0)F_k^{s\dagger}(y^0)
    \left( \begin{array}{cc}
     e_{\mathbf{k}}^{s\dagger} & \\ & e_{\mathbf{k}}^{s\dagger}\bar{\sigma}^0 \end{array} \right),
    \nonumber \\ \label{eq:Green_func}
\end{eqnarray}
where
\begin{equation}
 F_k^s\equiv \left( \begin{array}{cc}
  u_k^s & -v_k^{s*} \\ v_k^s & u_k^{s*} \end{array} \right),
\end{equation}
is a (retarded) Green function\footnote{In (\ref{eq:Green_func}), $\theta$-function means not a field, but a step function.}
from $y^0=t^{\rm in}$ to $x^0$.
Using a notation
\begin{equation}
 \tilde{G}_{xy}\equiv G_{xy}^{\rm in}\cdot\frac{1}{2}\dot{\theta}(y^0)
  \left( \begin{array}{cc} -\bar{\sigma}^0 & \\ & \sigma^0 \end{array} \right),
\end{equation}
then we can obtain a perturbative solution of (\ref{eq:YF_eq}) as
\begin{eqnarray}
 \left( \begin{array}{c} \xi_E(x) \\ \xi_E^\dagger(x) \end{array} \right)
  &=& \left( \begin{array}{c} \xi_E^{\rm in}(x) \\ \xi_E^{{\rm in}\dagger}(x) \end{array} \right)
   - \int d^4y \: \tilde{G}_{xy} \left(
   \begin{array}{c} \xi_E(y) \\ \xi_E^\dagger(y) \end{array} \right) \nonumber \\
  &=& \int d^4y [1+\tilde{G}]_{xy}^{-1}
   \left( \begin{array}{c} \xi_E^{\rm in}(y) \\ \xi_E^{{\rm in}\dagger}(y) \end{array} \right) \nonumber \\
  &=& \int d^4y \left(\delta^4(x-y)-\tilde{G}_{xy}\right.\nonumber \\
  & & \quad \left.+\int d^4z \tilde{G}_{xz}\tilde{G}_{zy}+\cdots\right)
   \left( \begin{array}{c} \xi_E^{\rm in}(y) \\ \xi_E^{{\rm in}\dagger}(y) \end{array} \right). \nonumber \\ \label{eq:YF_expand}
\end{eqnarray}

\subsection{Net number density}
The number density {\it operator} for each helicity can be shown as
\begin{eqnarray}
 \hat{n}^s &=& \int\frac{d^3k}{(2\pi)^3} 
  \left[ \left(1-\frac{s|\mathbf{k}|}{\omega_k}\right)\tilde{\Xi}_{\mathbf{k}}^{s\dagger}\tilde{\Xi}_{\mathbf{k}}^s
  +\left(1+\frac{s|\mathbf{k}|}{\omega_k}\right)\tilde{\Xi}_{\mathbf{-k}}^s\tilde{\Xi}_{\mathbf{-k}}^{s\dagger}\right.\nonumber \\
  & & \left. \qquad -\frac{|m|}{\omega_k}\left(se^{+i\rho_{\mathbf{k}}}\tilde{\Xi}_{\mathbf{-k}}^s\tilde{\Xi}_{\mathbf{k}}^s
   + \cdot se^{-i\rho_{\mathbf{k}}}\tilde{\Xi}_{\mathbf{k}}^{s\dagger}\tilde{\Xi}_{\mathbf{-k}}^{s\dagger}\right)\right]
   \label{eq:number_density_s}
\end{eqnarray}
where $\tilde{\Xi}_{\mathbf{k}}^s$ is a Fourier transformed operator of $\xi_E(x)$ defined by
\begin{equation}
 \xi_E=\int\frac{d^3k}{(2\pi)^3}\sum_se_{\mathbf{k}}^s\tilde{\Xi}_{\mathbf{k}}^s. \label{eq:Fourier_full_xi}
\end{equation}
Note that the representation of (\ref{eq:number_density_s}) is corresponding to [(kinetic energy) $-$ (vacuum energy)]/(1 particle energy).

The net number density of the Majorana fermion can be defined by the difference of helicity.
Using (\ref{eq:number_density_s}) and (\ref{eq:Fourier_full_xi}),
the net number density by the field operator $\xi_E$ is given by
\begin{eqnarray}
 & & n_{\rm net} \: \equiv \: n^+-n^- \\
 & & = \frac{1}{V}\int d^3xd^3y\frac{d^3k}{(2\pi)^3} e^{i\mathbf{k\cdot(x-y)}} \nonumber \\
 & & \quad \times \frac{1}{2}
  \langle0|\left[ -\xi_E^\dagger(t,\mathbf{x})\left(\frac{|\mathbf{k}|}{\omega_k}\bar{\sigma}^0
  +\frac{k^i}{|\mathbf{k}|}\bar{\sigma}^i\right)\xi_E(t,\mathbf{y}) \right. \nonumber \\
 & & \qquad \qquad +\xi_E(t,\mathbf{x})\left(\frac{|\mathbf{k}|}{\omega_k}\sigma^0
  +\frac{k^i}{|\mathbf{k}|}\sigma^i\right)\xi_E^\dagger(t,\mathbf{y}) \nonumber \\
 & & \qquad \qquad - \frac{|m_\xi|}{\omega_k}
  \xi_E(t,\mathbf{x})\frac{k^i}{|\mathbf{k}|}\sigma^0\bar{\sigma}^i\xi_E(t,\mathbf{y}) \nonumber \\
 & & \left. \qquad \qquad + \frac{|m_\xi|}{\omega_k}
  \xi_E^\dagger(t,\mathbf{x})\frac{k^i}{|\mathbf{k}|}\bar{\sigma}^0\sigma^i\xi_E^\dagger(t,\mathbf{y})\right]|0\rangle \nonumber\\
 & & = \frac{1}{V}\int d^3xd^3y\frac{d^3k}{(2\pi)^3} e^{i\mathbf{k\cdot(x-y)}} \nonumber \\
 & & \quad \times \frac{1}{2}
  \langle0|\left( \begin{array}{cc}\xi_E^\dagger(t,\mathbf{x}) & \xi_E(t,\mathbf{x}) \end{array} \right) 
  S_k\left(
   \begin{array}{c} \xi_E(t,\mathbf{y}) \\ \xi_E^\dagger(t,\mathbf{y}) \end{array}\right)|0\rangle \nonumber\\ \label{eq:full_net_number}
\end{eqnarray}
where we used a notation
\begin{equation}
 S_k \equiv \left( \begin{array}{cc}
    -\frac{|\mathbf{k}|}{\omega_k}\bar{\sigma}^0-\frac{k^i}{|\mathbf{k}|}\bar{\sigma}^i & \frac{|m_\xi|}{\omega_k}\frac{k^i}{|\mathbf{k}|}\bar{\sigma}^0\sigma^i \\
    -\frac{|m_\xi|}{\omega_k}\frac{k^i}{|\mathbf{k}|}\sigma^0\bar{\sigma}^i & \frac{|\mathbf{k}|}{\omega_k}\sigma^0+\frac{k^i}{|\mathbf{k}|}\sigma^i
   \end{array} \right).
\end{equation}
Substituting (\ref{eq:YF_expand}) into the above result, then
one can evaluate the net number density perturbatively in order of the Green function as
\begin{eqnarray}
 n_{\rm net} &=& \frac{1}{V}\int d^3xd^3y\frac{d^3k}{(2\pi)^3} e^{i\mathbf{k\cdot(x-y)}} \int d^4zd^4w\nonumber \\
 & & \quad \times \frac{1}{2}
  \langle0|\left( \begin{array}{cc}\xi_E^{{\rm in}\dagger}(z) & \xi_E^{\rm in}(z) \end{array} \right) \nonumber \\
 & & \qquad \qquad \times ([1+\tilde{G}]^{-1}_{xz})^\dagger S_k [1+\tilde{G}]^{-1}_{yw} \nonumber \\
 & & \qquad \qquad \times \left(\begin{array}{c} \xi_E^{\rm in}(w) \\ \xi_E^{{\rm in}\dagger}(w) \end{array}\right)|0\rangle_{x^0=y^0=t}\\
 &=& \frac{1}{V}\int d^3xd^3y\frac{d^3k}{(2\pi)^3} e^{i\mathbf{k\cdot(x-y)}} \int d^4zd^4w\nonumber \\
 & & \quad \times \frac{1}{2}
  \langle0|\Psi_z^\dagger \nonumber \\
 & & \qquad \quad \times \left[({\rm 0th})+({\rm 1st})+({\rm 2nd})+({\rm 3rd})+\cdots\right] \nonumber \\
 & & \qquad \quad \times \Psi_w|0\rangle_{x^0=y^0=t} \label{eq:full_net_number2}
\end{eqnarray}
where
\begin{eqnarray}
 \Psi_z &\equiv& \left(\begin{array}{c} \xi_E^{\rm in}(z) \\ \xi_E^{{\rm in}\dagger}(z) \end{array}\right),\\
 ({\rm 0th}) & \equiv & \delta^4(x-z)S_k\delta^4(y-w), \label{eq:s_zeroth}\\
 ({\rm 1st}) & \equiv & -([\tilde{G}]_{xz})^\dagger S_k\delta^4(y-w) -\delta^4(x-z)S_k\tilde{G}_{yw}\nonumber \\ \label{eq:s_first}\\
 ({\rm 2nd}) & \equiv & ([\tilde{G}\tilde{G}]_{xz})^\dagger S_k\delta^4(y-w)
  +(\tilde{G}_{xz})^\dagger S_k \tilde{G}_{yw}\nonumber \\
  & & +\delta^4(x-z)S_k[\tilde{G}\tilde{G}]_{yw}\nonumber \\ \label{eq:s_second}\\
 ({\rm 3rd}) & \equiv & -([\tilde{G}\tilde{G}\tilde{G}]_{xz})^\dagger S_k\delta^4(y-w)
  -([\tilde{G}\tilde{G}]_{xz})^\dagger S_k \tilde{G}_{yw}\nonumber \\
  & & -(\tilde{G}_{xz})^\dagger S_k [\tilde{G}\tilde{G}]_{yw}-\delta^4(x-z)S_k[\tilde{G}\tilde{G}\tilde{G}]_{yw}.\nonumber \\ \label{eq:s_third}
\end{eqnarray}

In the following, we will show that the contributions from the zeroth, the
first and the second order vanishes and the leading contribution appears
from the third order.

\subsubsection{0th order}
At the zeroth order, we just consider with (\ref{eq:s_zeroth}) in (\ref{eq:full_net_number2}). Then
\begin{eqnarray}
 & & (n_{\rm net})_{0{\rm th}} \\
 & & = \frac{1}{V}\int d^3xd^3y\frac{d^3k}{(2\pi)^3} e^{i\mathbf{k\cdot(x-y)}} \cdot \frac{1}{2}
  \langle0|\Psi_x^\dagger  S_k \Psi_y|0\rangle_{x^0=y^0=t} \nonumber \\
 & & = \int\frac{d^3k}{(2\pi)^3}\sum_s\frac{1}{2}\left( \begin{array}{cc} v_k^s(t) & u_k^s(t) \end{array} \right)
  \left( \begin{array}{cc} e_{\mathbf{k}}^{s\dagger} & \\ & -e_{\mathbf{k}}^{s\dagger}\bar{\sigma}^0 \end{array} \right) \nonumber \\
 & & \qquad \times S_k \left( \begin{array}{cc} e_{\mathbf{k}}^s & \\ & -\bar{\sigma}^0e_{\mathbf{k}}^s \end{array} \right)
  \left( \begin{array}{c} v_k^{s*}(t) \\ u_k^{s*}(t) \end{array} \right). \nonumber \\
 & & = \int\frac{d^3k}{(2\pi)^3}\sum_s\frac{1}{2}\left( \begin{array}{cc} v_k^s(t) & u_k^s(t) \end{array} \right) \nonumber \\
 & & \qquad \times s \left( \begin{array}{cc} 1-\frac{s|\mathbf{k}|}{\omega_k} & \frac{|m_\xi|}{\omega_k} \\
  -\frac{|m_\xi|}{\omega_k} & 1+\frac{s|\mathbf{k}|}{\omega_k} \end{array} \right)
  \left( \begin{array}{c} v_k^{s*}(t) \\ u_k^{s*}(t) \end{array} \right)\nonumber \\
 & & = \int\frac{d^3k}{(2\pi)^3}\sum_s\frac{s}{2}\left[ \left(1-\frac{s|\mathbf{k}|}{\omega_k}\right)|v_k^s|^2
  +\frac{|m_\xi|}{\omega_k}v_k^su_k^{s*} \right. \nonumber \\
 & & \qquad \left.-\frac{|m_\xi|}{\omega_k}u_k^sv_k^{s*}
  +\left(1+\frac{s|\mathbf{k}|}{\omega_k}\right)|u_k^s|^2 \right] \nonumber \\
 & & =0.
\end{eqnarray}

\subsubsection{1st order}
The first order of (\ref{eq:full_net_number2}) with (\ref{eq:s_first}) can be calculated as
\begin{eqnarray}
 & & (n_{\rm net})_{1{\rm st}} \nonumber \\
 & & = \frac{1}{V}\int d^3xd^3y\frac{d^3k}{(2\pi)^3} \int d^4z \nonumber \\
 & & \quad \times \frac{1}{2}
  \left[ - e^{i\mathbf{k\cdot(x-y)}} \langle0|\Psi_x^\dagger
   S_k \tilde{G}_{yz} \Psi_z |0\rangle_{x^0=y^0=t} +(H.c.) \right] \nonumber\\
 & & = \int\frac{d^3k}{(2\pi)^3} \left[ -\int_{t^{\rm in}}^tdt'\frac{i}{2}\dot{\theta}(t')\sum_s\frac{1}{2} \right.\nonumber \\
 & & \qquad \times \left( \begin{array}{cc} v_k^s(t) & u_k^s(t) \end{array} \right)
  \left( \begin{array}{cc} e_{\mathbf{k}}^{s\dagger} & \\ & -e_{\mathbf{k}}^{s\dagger}\bar{\sigma}^0 \end{array} \right) S_k \nonumber \\
 & & \qquad \times \left( \begin{array}{cc}
     e_{\mathbf{k}}^s & \\ & \bar{\sigma}^0e_{\mathbf{k}}^s \end{array} \right) F_k^s(t)F_k^{s\dagger}(t')
    \left( \begin{array}{cc}
     -e_{\mathbf{k}}^{s\dagger}\bar{\sigma}^0 & \\ & e_{\mathbf{k}}^{s\dagger} \end{array} \right) \nonumber \\
 & & \qquad \times \left. \left( \begin{array}{cc} e_{\mathbf{k}}^s & \\ & -\bar{\sigma}^0e_{\mathbf{k}}^s \end{array} \right)
  \left( \begin{array}{c} v_k^{s*}(t') \\ u_k^{s*}(t') \end{array} \right) + (H.c.) \right]. \nonumber \\ \nonumber\\
 & & = \int\frac{d^3k}{(2\pi)^3} \left[ -\int_{t^{\rm in}}^tdt'\frac{i}{2}\dot{\theta}(t')\sum_s\frac{1}{2}\right. \nonumber \\
 & & \qquad \times \left( \begin{array}{cc} v_k^s(t) & u_k^s(t) \end{array} \right)
  s\left( \begin{array}{cc} 1-\frac{s|\mathbf{k}|}{\omega_k} & \frac{|m_\xi|}{\omega_k} \\
  -\frac{|m_\xi|}{\omega_k} & -1-\frac{s|\mathbf{k}|}{\omega_k} \end{array} \right) F_k^s(t) \nonumber \\
 & & \qquad \left. \times F_k^{s\dagger}(t')\cdot(-1) \cdot
  \left( \begin{array}{c} v_k^{s*}(t') \\ u_k^{s*}(t') \end{array} \right) + (H.c.) \right].
\end{eqnarray}
Note that one can show
\begin{eqnarray}
 & & \left( \begin{array}{cc} v_k^s(t) & u_k^s(t) \end{array} \right)
  \left( \begin{array}{cc} 1-\frac{s|\mathbf{k}|}{\omega_k} & \frac{|m_\xi|}{\omega_k} \\
  -\frac{|m_\xi|}{\omega_k} & -1-\frac{s|\mathbf{k}|}{\omega_k} \end{array} \right) F_k^s(t) \nonumber \\
 & & = \left( \begin{array}{cc} 0 & 0 \end{array} \right), \label{eq:projection}
\end{eqnarray}
thus
\begin{equation}
 (n_{\rm net})_{1{\rm st}} =0.
\end{equation}
Especially, the fact (\ref{eq:projection}) indicates
\begin{equation}
 \langle0|\Psi_x^\dagger S_k G_{yz} \times \cdots |0\rangle =0 \label{eq:zero_term}
\end{equation}
and its Hermite conjugate is also same.
In order to obtain nonzero terms, we need additional Green function
between $\Psi_x^\dagger$ and $S_k$.

\subsubsection{2nd order}
Taking into account (\ref{eq:zero_term}), the survival term in (\ref{eq:s_second})
is only the second term $(\tilde{G}_{xz})^\dagger S_k\tilde{G}_{yz}$.
Thus the second order of (\ref{eq:full_net_number2}) is
\begin{eqnarray}
 & & (n_{\rm net})_{2{\rm nd}} \nonumber \\
 & & = \frac{1}{V}\int d^3xd^3y\frac{d^3k}{(2\pi)^3} \int d^4z_1d^4z_2\nonumber \\
 & & \quad \times \frac{1}{2}
  \langle0| e^{i\mathbf{k\cdot(x-y)}} \Psi_{z_1}^\dagger
  [\tilde{G}_{xz_1}]^\dagger S_k \tilde{G}_{yz_2} \Psi_{z_2} |0\rangle_{x^0=y^0=t}  \nonumber\\
 & & = \int\frac{d^3k}{(2\pi)^3} \int_{t^{\rm in}}^tdt_1\int_{t^{\rm in}}^tdt_2
  \frac{1}{4}\dot{\theta}(t_1)\dot{\theta}(t_2)\sum_s\frac{1}{2} \nonumber \\
 & & \quad \times \left( \begin{array}{cc} v_k^s(t_1) & u_k^s(t_1) \end{array} \right)
  \left( \begin{array}{cc} e_{\mathbf{k}}^{s\dagger} & \\ & -e_{\mathbf{k}}^{s\dagger}\bar{\sigma}^0 \end{array} \right) \nonumber \\
 & & \qquad \times \left( \begin{array}{cc}
     -\bar{\sigma}^0e_{\mathbf{k}}^s & \\ & e_{\mathbf{k}}^s \end{array} \right) F_k^s(t_1)F_k^{s\dagger}(t)
    \left( \begin{array}{cc}
     e_{\mathbf{k}}^{s\dagger} & \\ & e_{\mathbf{k}}^{s\dagger}\bar{\sigma}^0 \end{array} \right) \nonumber \\
 & & \qquad \times S_k \left( \begin{array}{cc}
     e_{\mathbf{k}}^s & \\ & \bar{\sigma}^0e_{\mathbf{k}}^s \end{array} \right) F_k^s(t)F_k^{s\dagger}(t_2)
    \left( \begin{array}{cc}
     -e_{\mathbf{k}}^{s\dagger}\bar{\sigma}^0 & \\ & e_{\mathbf{k}}^{s\dagger} \end{array} \right) \nonumber \\
 & & \qquad \times \left( \begin{array}{cc} e_{\mathbf{k}}^s & \\ & -\bar{\sigma}^0e_{\mathbf{k}}^s \end{array} \right)
  \left( \begin{array}{c} v_k^{s*}(t_2) \\ u_k^{s*}(t_2) \end{array} \right) \nonumber \\
 & & = \int\frac{d^3k}{(2\pi)^3}\int_{t^{\rm in}}^tdt_1\int_{t^{\rm in}}^tdt_2
  \frac{1}{4}\dot{\theta}(t_1)\dot{\theta}(t_2)\sum_s\frac{1}{2} \nonumber \\
 & & \quad \times \left( \begin{array}{cc} v_k^s(t_1) & u_k^s(t_1) \end{array} \right)F_k^s(t_1) \nonumber \\
 & & \qquad \times F_k^{s\dagger}(t)\cdot s \left( \begin{array}{cc}
    1-\frac{s|\mathbf{k}|}{\omega_k} & \frac{|m_\xi|}{\omega_k} \\
    \frac{|m_\xi|}{\omega_k} & 1+\frac{s|\mathbf{k}|}{\omega_k} \end{array} \right) F_k^s(t) \nonumber \\
 & & \qquad \times F_k^{s\dagger}(t_2)
    \left( \begin{array}{c} v_k^{s*}(t_2) \\ u_k^{s*}(t_2) \end{array} \right) \nonumber \\
 & & = \int\frac{d^3k}{(2\pi)^3}\int_{t^{\rm in}}^tdt_1\int_{t^{\rm in}}^tdt_2
  \frac{1}{4}\dot{\theta}(t_1)\dot{\theta}(t_2)\sum_s\frac{1}{2} \nonumber \\
 & & \quad \times \left( \begin{array}{cc} v_k^s(t_1) & u_k^s(t_1) \end{array} \right)F_k^s(t_1) \nonumber \\
 & & \qquad \times s \left( \begin{array}{cc} 2 & \\ & 0 \end{array} \right) \cdot F_k^{s\dagger}(t_2)
    \left( \begin{array}{c} v_k^{s*}(t_2) \\ u_k^{s*}(t_2) \end{array} \right) \nonumber \\
 & & = \int\frac{d^3k}{(2\pi)^3}\int_{t^{\rm in}}^tdt_1\int_{t^{\rm in}}^tdt_2
  \frac{1}{4}\dot{\theta}(t_1)\dot{\theta}(t_2) \nonumber \\
 & & \quad \times \sum_s s\cdot u_k^s(t_1) v_k^s(t_1) \cdot u_k^{s*}(t_2) v_k^{s*}(t_2) \nonumber \\
 & & = 0.
\end{eqnarray}
Therefore, the second order also cannot affect the asymmetry.

\subsubsection{3rd order}
Taking into account (\ref{eq:zero_term}), the survival terms in (\ref{eq:s_third})
are $([\tilde{G}\tilde{G}]_{xz})^\dagger S_k \tilde{G}_{yw}$.
Thus the contribution to the net number from the third order is
\begin{eqnarray}
 & & (n_{\rm net})_{3{\rm rd}} \nonumber \\
 & & = \frac{1}{V}\int d^3xd^3y\frac{d^3k}{(2\pi)^3} \int d^4z_1d^4z_2d^4z_3\nonumber \\
 & & \quad \times \frac{1}{2}
  \langle0| \left[ - e^{i\mathbf{k\cdot(x-y)}} \Psi_{z_1}^\dagger [G_{xz_1}]^\dagger \right. \nonumber \\
 & & \qquad \qquad \left. \times S_k G_{yz_2}G_{z_2z_3} \Psi_{z_3} + (H.c.) \right]|0\rangle \nonumber \\
 & & = \int\frac{d^3k}{(2\pi)^3} \left[ -\int_{t^{\rm in}}^tdt_1\int_{t^{\rm in}}^tdt_2\int_{t^{\rm in}}^{t_2}dt_3
  \frac{i}{8}\dot{\theta}(t_1)\dot{\theta}(t_2)\dot{\theta}(t_3) \right. \nonumber \\
 & & \quad \times \sum_s\frac{1}{2} \left( \begin{array}{cc} v_k^s(t_1) & u_k^s(t_1) \end{array} \right)
  \left( \begin{array}{cc} e_{\mathbf{k}}^{s\dagger} & \\ & -e_{\mathbf{k}}^{s\dagger}\bar{\sigma}^0 \end{array} \right) \nonumber \\
 & & \qquad \times \left( \begin{array}{cc}
     -\bar{\sigma}^0e_{\mathbf{k}}^s & \\ & e_{\mathbf{k}}^s \end{array} \right) F_k^s(t_1)F_k^{s\dagger}(t)
    \left( \begin{array}{cc}
     e_{\mathbf{k}}^{s\dagger} & \\ & e_{\mathbf{k}}^{s\dagger}\bar{\sigma}^0 \end{array} \right) \nonumber \\
 & & \qquad \times S_k \left( \begin{array}{cc}
     e_{\mathbf{k}}^s & \\ & \bar{\sigma}^0e_{\mathbf{k}}^s \end{array} \right) F_k^s(t)F_k^{s\dagger}(t_2)
    \left( \begin{array}{cc}
     -e_{\mathbf{k}}^{s\dagger}\bar{\sigma}^0 & \\ & e_{\mathbf{k}}^{s\dagger} \end{array} \right) \nonumber \\
 & & \qquad \times \left( \begin{array}{cc}
     e_{\mathbf{k}}^s & \\ & \bar{\sigma}^0e_{\mathbf{k}}^s \end{array} \right) F_k^s(t_2)F_k^{s\dagger}(t_3)
    \left( \begin{array}{cc}
     -e_{\mathbf{k}}^{s\dagger}\bar{\sigma}^0 & \\ & e_{\mathbf{k}}^{s\dagger} \end{array} \right) \nonumber \\
 & & \qquad \left. \times \left( \begin{array}{cc} e_{\mathbf{k}}^s & \\ & -\bar{\sigma}^0e_{\mathbf{k}}^s \end{array} \right)
  \left( \begin{array}{c} v_k^{s*}(t_3) \\ u_k^{s*}(t_3) \end{array} \right) + (H.c.) \right]\nonumber \\
 & & = \int\frac{d^3k}{(2\pi)^3} \left[ -\int_{t^{\rm in}}^tdt_1\int_{t^{\rm in}}^tdt_2\int_{t^{\rm in}}^{t_2}dt_3
    \frac{i}{8}\dot{\theta}(t_1)\dot{\theta}(t_2)\dot{\theta}(t_3) \right. \nonumber \\
 & & \quad \times \sum_s\frac{1}{2} \left( \begin{array}{cc} v_k^s(t_1) & u_k^s(t_1) \end{array} \right)F_k^s(t_1)
    \cdot s \left( \begin{array}{cc} 2 & \\ & 0 \end{array} \right) \nonumber \\
 & & \qquad \times F_k^{s\dagger}(t_2)
    \left( \begin{array}{cc} -1 & \\ & 1 \end{array} \right) F_k^s(t_2) \nonumber \\
 & & \qquad \left. \times  F_k^{s\dagger}(t_3) 
  \left( \begin{array}{c} v_k^{s*}(t_3) \\ u_k^{s*}(t_3) \end{array} \right) + (H.c.) \right]\nonumber \\
 & & = \int\frac{d^3k}{(2\pi)^3} \left[ -\int_{t^{\rm in}}^tdt_1\int_{t^{\rm in}}^tdt_2\int_{t^{\rm in}}^{t_2}dt_3
    \frac{i}{8}\dot{\theta}(t_1)\dot{\theta}(t_2)\dot{\theta}(t_3) \right. \nonumber \\
 & & \quad \times \sum_s s\cdot2u_k^s(t_1)v_k^s(t_1) \nonumber \\
 & & \qquad \times \left( -(|u_k^s(t_2)|^2-|v_k^s(t_2)|^2)\cdot2u_k^{s*}(t_3)v_k^{s*}(t_3) \right. \nonumber \\
 & & \qquad \left. \left. +2u_k^{s*}(t_2)v_k^{s*}(t_2)\cdot(|u_k^s(t_3)|^2-|v_k^s(t_3)|^2) \right) + (H.c.) \right]\nonumber \\
 & & = \int\frac{d^3k}{(2\pi)^3} \int_{t^{\rm in}}^tdt_1\int_{t^{\rm in}}^tdt_2\int_{t^{\rm in}}^{t_2}dt_3
    \frac{i}{8}\dot{\theta}(t_1)\dot{\theta}(t_2)\dot{\theta}(t_3) \nonumber \\
 & & \quad \times 2 \frac{|\mathbf{k}|}{\omega_k}\frac{|m_\xi|^2}{\omega_k^2}e^{-2i\omega_kt_1}
    \left( e^{+2i\omega_kt_2}-e^{+2i\omega_kt_3} \right) + (H.c.) \nonumber \label{eq:third_oder_result} \\
\end{eqnarray}
In order to evaluate the time integral, we choose $t^{\rm in}=0$
and assume the initial and final condition of $\theta(t)$ as
\begin{equation}
 \theta(t^{\rm in})\equiv\theta_i, \quad \theta(t\rightarrow\infty)=0,
\end{equation}
\begin{equation}
 \dot{\theta}(t^{\rm in})=\dot{\theta}(t\rightarrow\infty)=0.
\end{equation}
Then each time integral with (\ref{eq:eom_theta_approx}) gives
\begin{eqnarray}
 & & \int_0^t dt' \dot{\theta}(t') e^{+2i\omega t'} \nonumber  \\
 & & = -\frac{1}{4\omega_k^2-m_\theta^2+2i\omega_k\Gamma}\nonumber \\
 & & \quad \times \left[2i\omega_k\dot{\theta}(t)e^{2i\omega t}
  + m_\theta^2\left(\theta(t)e^{2i\omega t}-\theta_i\right) \right],
\end{eqnarray}
\begin{eqnarray}
 & & \int_0^\infty dt' \dot{\theta}^2(t') e^{+2i\omega t'} \nonumber  \\
 & & = \frac{m_\theta^4}{2(2i\omega_k-\Gamma)(\omega_k^2-m_\theta^2+i\omega_k\Gamma)}
  \theta_i^2,
\end{eqnarray}
\begin{eqnarray}
 & & \int_0^\infty dt' \theta(t')\dot{\theta}(t') e^{+2i\omega t'} \nonumber  \\
 & & = \frac{m_\theta^2(i\omega_k-\Gamma)}{2(2i\omega_k-\Gamma)(\omega_k^2-m_\theta^2+i\omega_k\Gamma)}
  \theta_i^2.
\end{eqnarray}
Applying these formulae to (\ref{eq:third_oder_result}), finally we can obtain
\begin{eqnarray}
 (n_{\rm net})_{3{\rm rd}}
  &=& \int\frac{d^3k}{(2\pi)^3}\frac{|\mathbf{k}|}{\omega_k}
   \frac{|m_\xi|^2}{\omega_k^2}\frac{1}{2}\omega_km_\theta^6\Gamma\theta_i^3 \nonumber \\
  & &\quad \times \frac{1}{(4\omega_k^2-m_\theta^2)^2+4\omega_k^2\Gamma^2} \nonumber \\
  & & \quad \times \frac{1}{(\omega_k^2-m_\theta^2)^2+\omega_k^2\Gamma^2}\frac{7\omega_k^2-m_\theta^2+\Gamma^2}{\omega_k^2+\Gamma^2}
   \nonumber \\
  &\sim& \frac{1}{4\pi}|m_\xi|^2m_\theta\theta_i^3,
\end{eqnarray}
where we used an assumption $m_\theta \gg |m_\xi| \gg \Gamma$ and
the narrow width approximation
\begin{equation}
 \frac{1}{(\omega^2-m_\theta^2)^2+m_\theta^2\Gamma^2}\sim\frac{\pi}{m_\theta\Gamma}\delta(\omega^2-m_\theta^2). 
\end{equation}
Thus the nonzero term of the mean net number appears at the third order.

\section{Bias in the eigenstates}
\label{app-bias}
The asymmetry can be seen as the biased mixing in the
eigenstates.
In this section, we are going to review basic strategies in this direction.

\subsection{Quantum corrections}
It would be useful to remember how the bias appears from the loop corrections.
Although many fields are required for the quantum correction, which is
not explicitly discussed in this paper, the correction is symbolically
given by a Hermite matrix $\Gamma$, which  
modifies the Hamiltonian as $H\rightarrow H-i\Gamma$.
As the result, the correction to the Hamiltonian is anti-Hermite ($\sim
i\Gamma$), where $\Gamma$ is given by
\begin{eqnarray}
\Gamma&=& 
\left(
\begin{array}{cc}
\Gamma_d & \Gamma_\Delta\\
(\Gamma_{\Delta})^*& \Gamma_d
\end{array}
\right).
\end{eqnarray}
Note that the correction ($i\Gamma$) is (1) anti-Hermite, and  (2) the
imaginary part of the off-diagonal elements ($\Im\left[\Gamma_\Delta\right])$ 
are important for the bias.
The real part of $\Gamma_\Delta$ may appear at the
tree level, but it does not cause the matter-antimatter asymmetry.
To see the matter-antimatter asymmetry (bias) in the eigenstates,
it is useful to calculate eigenvectors of the matrix given by
\begin{eqnarray}
\left(
\begin{array}{cc}
H_{11} & H_{12}\\
H_{21} & H_{21}
\end{array}
\right)&=& 
\left(
\begin{array}{cc}
M& \Delta-i\Gamma_\Delta\\
\Delta -i\Gamma_\Delta^*& M
\end{array}
\right).
\end{eqnarray}
Here the diagonal element $\Gamma_d$ is absorbed into $M$ for simplicity.
For this model, the eigenvectors are written as
\begin{eqnarray}
\left(\pm\frac{r}{\sqrt{1+|r|^2}},\frac{1}{\sqrt{1+|r|^2}}\right),
\end{eqnarray}
where $r\equiv
\sqrt{\frac{\Delta-i\Gamma_\Delta}{\Delta^*-i\Gamma_\Delta^*}}$.
The eigenvectors are biased when $|r|\ne 1$.
This parameter is commonly used to measure the CP violation in a
kaon. 
The biases are the same for both eigenstates,
which suggests that these eigenstates are generating the same bias.
This is the crucial difference from the chemical potential.
The eigenvalues are
$M\pm\sqrt{(\Delta-i\Gamma_\Delta)(\Delta-i\Gamma_\Delta^*)}$.
Note that the above arguments are based on the system with the kaon, where
many other fields are implicitly assumed to generate the quantum correction.

\subsection{The Chemical potential
(in the limit of $\dot{\mu}\simeq 0$)}
Similar bias can be introduced by the chemical potential.
Of course, the statistical bias is obvious in the thermal background 
(as far as the chemical potential appears in the Hamiltonian),
but the thermal equilibrium is not considered here.
Here, the particle production is assumed to be nonthermal, and the
decay of the particle is assumed to be much slower than the mixings.
Initially, the chemical potential is supposed to be a constant
(i.e, the Berry connection gives a Hermite and time-independent
contribution).
More simply, one may think that the particle production proceeds with
the pure eigenstates.
Later in Sec. \ref{sec-decay}, we will consider the opposite case, in
which the particle production cannot be explained by the
eigenstates.
 
At the beginning of this section, we have explained that
$\dot{\theta}\ne 0$ can introduce the correction, which becomes
\begin{eqnarray}
i\frac{d}{dt}\psi^R&=&(H^R -\gamma_G)\psi^R,\nonumber\\
H^R&=& 
\left(
\begin{array}{cc}
M & |\Delta|\\
|\Delta| & M
\end{array}
\right),\nonumber\\
\gamma_G&=&
\left(
\begin{array}{cc}
\mu & 0\\
0 & -\mu
\end{array}
\right).
\end{eqnarray}
Unlike the quantum correction discussed above, the geometric correction
caused by $\dot{\theta}\ne 0$ is (1) Hermite, and (2) diagonal elements
determine the bias.
Here we are comparing $i\Gamma$ with $\gamma_G$.

To see the matter-antimatter asymmetry (bias) in the eigenstates,
it is useful to calculate eigenvectors of the matrix given by
\begin{eqnarray}
\left(
\begin{array}{cc}
H_{11} & H_{12}\\
H_{21} & H_{21}
\end{array}
\right)&=& 
\left(
\begin{array}{cc}
M+\mu& |\Delta|\\
|\Delta|& M-\mu
\end{array}
\right).
\end{eqnarray}
Choosing a parameter $p\equiv \mu/|\Delta|$, 
the eigenvalues are written as $M\pm |\Delta|\sqrt{1+p^2}$,
and their eigenvectors are given by
\begin{eqnarray}
\left(\pm\sqrt{\sqrt{1+p^2}\pm p},\sqrt{\sqrt{1+p^2}\mp p}\right).
\end{eqnarray}
One can see that the matter-antimatter bias is appearing in the
eigenstates, but there are significant
differences from the quantum corrections.

\subsection{Majorana Fermions}
\label{sec:majorana}
The above arguments for the matter-antimatter bias can be
applied to the Majorana fermions. 
In the past, the one-loop correction for the wave function mixing of singlet
(Majorana) neutrinos has been calculated in
Refs.\cite{Covi:1996wh,Flanz:1996fb, Rangarajan:1999kt}. 
We are not reviewing the calculation here, but the result is quite 
similar to the kaon quantum correction we have mentioned above.
For the quantum correction, the bias in the eigenstates appears in the
same manner as the bosonic field. 

Let us consider the bias caused by the chemical potential.
Our calculation for the chemical potential can be applied
straightforwardly to the Majorana fermion. 
We introduce the Majorana mass given by
\begin{eqnarray}
{\cal L}_m&=&
\left(m_R\bar{\psi_L^c}\psi_R+m_R^*\bar{\psi_R}\psi_L^c\right),
\end{eqnarray}
where $m_R$ is the Majorana mass for the singlet Right-handed 
Fermion.
Note that in our notation, $(\psi_R)^c=\psi_L^c$ and $(\psi_L)^c=\psi_R^c$.
Using $\Psi^t_R\equiv(\psi_R,\psi_L^c)$, one can write
\begin{eqnarray}
{\cal L}_m&=&\bar{\Psi}_R
\left(
\begin{array}{cc}
   0   & m_R \\
   m_R^*   & 0 
\end{array}
\right)\Psi_R.
\end{eqnarray}
For a static and homogeneous background, the eigenstates are
$\psi^E_\pm=\psi_R\pm\psi_L^c$, which satisfies
the Majorana condition $(\psi^E_\pm)^c=\pm \psi^E_\pm$.
On the other hand, if the Majorana mass is given by
$m_R=|m_R|e^{i\theta(t)}$, the Berry connection, which gives the
effective chemical potential, introduces mixing between the eigenstates,
and thus the ``eigenstates'' $\psi^E_\pm$ are no longer the true eigenstates.
Now the situation is the same as the bosonic scenario.
One can choose $m_R=|m_R|e^{i\theta(t)}$ to find the chemical potential
$\mu\equiv\dot{\theta}$ in the matrix,
\begin{eqnarray}
{\cal L}_m&=&\bar{\Psi}_R
\left(
\begin{array}{cc}
   \mu   & m_R \\
   m_R^*   & -\mu 
\end{array}
\right)\Psi_R.
\end{eqnarray}
On the other hand, if one considers the ``eigenstates'' of the mass matrix, one will find
\begin{eqnarray}
{\cal L}_m&=&\bar{\Psi}^E
\left(
\begin{array}{cc}
   |m_R|   & 0 \\
   0   & -|m_R|
\end{array}
\right)\Psi^E
\end{eqnarray}
and the Berry connection 
\begin{eqnarray}
iU^{-1}\dot{U}&=&
\frac{1}{2}\left(
\begin{array}{cc}
 & \dot{\theta}\\
\dot{\theta} &
\end{array}
\right).
\end{eqnarray}

Since we are choosing $\Psi^R$ for the calculation, 
the eigenvalues of the matrix are written as $\pm |m_R|\sqrt{1+p^2}$,
where we defined $p\equiv \mu/|m_R|$, 
 and their eigenvectors are given by
\begin{eqnarray}
\left(p\pm\sqrt{1+p^2},1\right),
\end{eqnarray}
which is biased when $p\ne 0$.
In this case, however, the bias in the eigenstates may cancel with each
other.
This cancellation is exact when (1) $\dot{\mu}= 0$ and
(2) the production and the decay rates are
indistinguishable. (Note that bosons are distinguishable when $M\ne 0$
and $\Delta \ne 0$.)
On the other hand, since the mixing rates of the CP-even and CP-odd states are
different, one can expect that the decay rates of these
eigenstates could be different in a phenomenological
situation.\footnote{The most useful example in this direction is the kaon.}
Moreover, these ``eigenstates'' are not the true eigenstates
when $\dot{\mu}\ne 0$.

One can introduce the Dirac mass $m_D$ to extend the model.
Temporarily, we assume that $|m_R|$ is constant but
$m_R$ is given by $m_R=|m_R|e^{i2\mu t}$.
Using $\Psi^t_0\equiv(\psi_L,\psi_R^c,\psi_R,\psi_L^c)$, one can write
\begin{eqnarray}
{\cal L}_m&=&\bar{\Psi}_0
\left(
\begin{array}{cccc}
 0   & 0 & m_D & 0 \\
 0 & 0   & 0   & m_D \\
 m_D   & 0   & 0   & m_R \\
 0 &  m_D  & m_R^*   & 0 
\end{array}
\right)\Psi_0.
\end{eqnarray}
Then, we find the chemical potential,
\begin{eqnarray}
\label{mass-fermi}
{\cal L}_m&=&\bar{\Psi}^R
\left(
\begin{array}{cccc}
 \mu   & 0 & m_D & 0 \\
 0 & -\mu   & 0   & m_D \\
 m_D   & 0   & \mu   & |m_R| \\
 0 &  m_D  & |m_R|   & -\mu 
\end{array}
\right)\Psi^R,
\end{eqnarray}
where $\Psi_R =U_\theta \Psi_0$ and 
$U_\theta={\rm diag}(e^{i \mu t},e^{-i\mu t},e^{i \mu t},e^{-i\mu t})$. 
If the Dirac mass is absent ($m_D=0$), one will recover the simplest scenario
\begin{eqnarray}
\label{mass-fermi2}
{\cal L}_m&=&\bar{\Psi}^R
\left(
\begin{array}{cccc}
 0   & 0 & 0 & 0 \\
 0 & 0   & 0   & 0 \\
 0  & 0   & \mu   & |m_R| \\
 0 &  0  & |m_R|   & -\mu 
\end{array}
\right)\Psi^R,
\end{eqnarray}
where $U_\theta$ is $U_\theta={\rm diag}(0,0,e^{i \mu t},e^{-i\mu t})$.

If $|m_R|$ is not constant but varies with time, 
the eigenstates can be written as 
\begin{eqnarray}
\Psi^E&=&U_1^{-1} \Psi^R = U_1^{-1} U_\theta^{-1} \Psi_0 \equiv
 U^{-1} \Psi_0, 
\end{eqnarray}
where not only $U_\theta$ but also $U_1$ could be time dependent.
Then for the ``eigenstate'' $\Psi^E$, the nonadiabatic Berry phase 
gives the contribution
\begin{eqnarray}
iU^{-1}\dot{U}&=& iU_1^{-1}(U_\theta^{-1}\dot{U}_\theta )U_1
+iU_1^{-1}\dot{U}_1,
\end{eqnarray}
which causes mixing between eigenstates.

Note that the above arguments are not valid when the chemical potential
is changing fast.

\end{document}